\documentclass[
amsmath,amssymb,
prd,nofootinbib,floatfix,12pt,
]{revtex4}

\newcommand\beq{\begin{eqnarray}}
\newcommand\eeq{\end{eqnarray}}
\def\lsim{\mathrel{\rlap{\lower4pt\hbox{$\sim$}}
    \raise1pt\hbox{$<$}}}                
\def\gsim{\mathrel{\rlap{\lower4pt\hbox{$\sim$}}
    \raise1pt\hbox{$>$}}}            

\allowdisplaybreaks
\usepackage{graphicx}
\usepackage{setspace}
\usepackage{accents}

\DeclareMathSymbol{\widehatsym}{\mathord}{largesymbols}{"62}
\newcommand\lowerwidehatsym{%
  \text{\smash{\raisebox{-1.3ex}{%
    $\widehatsym$}}}}
\newcommand\fixwidehat[1]{%
  \mathchoice
    {\accentset{\displaystyle\lowerwidehatsym}{#1}}
    {\accentset{\textstyle\lowerwidehatsym}{#1}}
    {\accentset{\scriptstyle\lowerwidehatsym}{#1}}
    {\accentset{\scriptscriptstyle\lowerwidehatsym}{#1}}
}

\DeclareMathSymbol{\widetildesym}{\mathord}{largesymbols}{"65}
\newcommand\lowerwidetildesym{%
  \text{\smash{\raisebox{-1.3ex}{%
    $\widetildesym$}}}}
\newcommand\fixwidetilde[1]{%
  \mathchoice
    {\accentset{\displaystyle\lowerwidetildesym}{#1}}
    {\accentset{\textstyle\lowerwidetildesym}{#1}}
    {\accentset{\scriptstyle\lowerwidetildesym}{#1}}
    {\accentset{\scriptscriptstyle\lowerwidetildesym}{#1}}
}

\newcommand\MSSM{\mbox{\scriptsize MSSM}}

\begin{document}
\renewcommand{\theequation}{\arabic{section}.\arabic{equation}}
\renewcommand{\thefigure}{\arabic{section}.\arabic{figure}}
\renewcommand{\thetable}{\arabic{section}.\arabic{table}}

\title{\large \baselineskip=20pt 
Quasi-fixed points from scalar sequestering and \\
the little hierarchy problem in supersymmetry}

\author{Stephen P.~Martin}
\affiliation{{\it Department of Physics, Northern Illinois University, DeKalb IL 60115}}

\begin{abstract}\normalsize \baselineskip=16pt 
In supersymmetric models with scalar sequestering, 
superconformal strong dynamics in the hidden sector suppresses the low-energy
couplings of mass dimension two, compared to the squares of the dimension 
one parameters. Taking into account restrictions on the anomalous dimensions in
superconformal theories, I point out that the interplay between the
hidden and visible sector renormalizations gives rise to 
quasi-fixed point running for the supersymmetric Standard Model
squared mass parameters, rather than  driving them to 0.
The extent to which this dynamics can ameliorate the little hierarchy problem
in supersymmetry is studied. Models of this type in which the gaugino
masses do not unify are arguably more natural, and are certainly
more likely to be accessible, eventually, to the Large Hadron Collider.
\end{abstract}

\maketitle

\vspace{-0.75cm}

\tableofcontents

\baselineskip=16.6pt

\setcounter{footnote}{1}
\setcounter{figure}{0}
\setcounter{table}{0}

\vspace{-0.15cm}


\section{Introduction\label{sec:intro}}
\setcounter{equation}{0}
\setcounter{figure}{0}
\setcounter{table}{0}
\setcounter{footnote}{1}

Low-energy supersymmetry \cite{primer} has historically been one of the 
most well-studied solutions for the hierarchy problem associated 
with the electroweak scale. This popularity has been on the wane as the 
continuing explorations of the Large Hadron Collider (LHC) have so far 
not found any evidence for the existence of superpartners. However, it 
is notable that no evidence for any of the other proposed solutions of 
the hierarchy problem has been found either; LHC searches for 
new physics have not produced any enduring positive signals. This state of 
affairs suggests that, regardless of the fate of supersymmetry, 
some new idea might be needed in order to understand the small size of 
the electroweak scale.

Within the context of supersymmetry, the problem is sometimes called the 
``little hierarchy problem", and can be illustrated with an equation 
that relates electroweak symmetry breaking to the parameters of the Minimal 
Supersymmetric Standard Model (MSSM) Lagrangian:
\beq
-\frac{1}{2} m_Z^2 &=& m^2_{H_u} + |\mu|^2 + 
\frac{1}{2v_u} \frac{\partial(\Delta V)}{\partial v_u} 
+ {\cal O}(1/\tan^2\beta) .
\label{eq:MZhierarchy}
\eeq 
This equation follows from minimizing the Higgs potential, and 
relates the $Z$ boson mass to the supersymmetry-preserving and -breaking 
Higgs squared masses $|\mu|^2$ and $m^2_{H_u}$ (in the notation of 
\cite{primer}) and the loop-suppressed corrections to the effective 
potential, $\Delta V$, which depends on a vacuum expectation value $v_u$ 
for the Higgs field
that couples to the up-type quarks and squarks. 
These loop corrections can be made small by 
an appropriate choice of renormalization scale, typically of order the geometric
mean of the top-squark masses. The remaining tree-level and loop 
contributions in eq.~(\ref{eq:MZhierarchy}), suppressed by 
$1/\tan^2\beta$ for large $\tan\beta$, 
are small enough to be neglected in a first 
approximation, if $\tan\beta$ is big as indicated by the observed Higgs 
scalar boson mass of 125 GeV. The little hierarchy problem is that in the MSSM, 
boundary conditions and 
radiative corrections correlate $m^2_{H_u}$ to the mass scales of the 
superpartners that now seem to be much heavier than $m_Z$, with lower mass bounds 
that continue to rise with each new reported LHC search, and 
top-squark and other superpartner masses well above 1 TeV also favored independently
by $M_h = 125$ GeV.
 
There is no hierarchy problem associated with $\mu$, which is a 
superpotential parameter and therefore protected by a chiral symmetry; 
the smallness of its magnitude compared to any larger mass 
scale is technically 
natural. This has lead to considerations of ``natural supersymmetry" 
scenarios, which in general suppose 
that somehow (with the complete explanation perhaps 
postponed) $m^2_{H_u}$ is comparable to $-|\mu|^2$, and 
both are not larger than the square of a few hundred GeV, so that the 
observed value of $m_Z$ could ensue without too 
much\footnote{Fine-tuning is an inherently fuzzy criterion. Therefore I 
make no further attempt to quantify it, as it is not possible to do 
so in a purely scientific way. Nevertheless, it is certainly useful, 
and even necessary, for scientists as a personal and subjective guide for
deciding how to allocate scarce resources such as time and 
money. The practical meaning of the words ``too much" is therefore left 
to the reader.} fine-tuning. The most prominent feature of the so-called 
natural supersymmetry scenario is that since $|\mu|$ should not be too 
large, the Higgsinos should be relatively light. 

Of course, the idea 
that natural supersymmetry requires small $|\mu|$ should be examined 
critically. Some proposals that seek to decouple the Higgsino masses 
from the supersymmetric little hierarchy problem in various ways have 
appeared in 
refs.~\cite{Dimopoulos:2014aua,Cohen:2015ala,Nelson:2015cea,Martin:2015eca}. 
The present paper is motivated by the possibility that the little 
hierarchy problem can be ameliorated in another way that does not 
require small $|\mu|$, by finding a reason why the particular 
combination
\beq
m^2_{H_u} + |\mu|^2,
\eeq
which appears as the non-loop-suppressed part of eq.~(\ref{eq:MZhierarchy}), 
can be dynamically driven towards small values, even if the 
individual terms in it are not, and even if all superpartner squared masses 
are much larger. 

One possible approach to realization of this comes from conformal sequestering 
\cite{Luty:2001jh,Luty:2001zv,Dine:2004dv,Ibe:2005pj,
Ibe:2005qv,Schmaltz:2006qs,Cohen:2006qc,Kachru:2007xp}, 
the proposal that supersymmetry breaking 
effects in the visible (MSSM) sector are renormalized by strong and 
nearly conformal dynamics in a hidden sector related to supersymmetry 
breaking. In a refinement of this idea, called scalar sequestering 
\cite{Roy:2007nz,Murayama:2007ge,Perez:2008ng}, the scalar squared 
masses have an extra suppression, compared to the dimension one 
parameters, due to the strong dynamics. This has the virtue of also 
ameliorating flavor violation problems that can arise in supersymmetry due to sfermion mixing. 
Other works that build on these ideas can be found in 
\cite{Kawamura:2008bf,Asano:2008qc,Kim:2009sy,Craig:2009rk,Hanaki:2010xf,
Craig:2013wga,Knapen:2013zla}, and the 
idea that conformal dynamics coupled directly to the MSSM sector 
can suppress flavor violation has been proposed in 
refs.~\cite{Nelson:2000sn,Nelson:2001mq}. 
From the point of view of the present paper, it is particularly intriguing 
that in theories of scalar sequestering, as pointed out in 
refs.~\cite{Murayama:2007ge,Perez:2008ng},
the dimension two parameters that undergo conformal scaling 
include the combined quantity 
$m^2_{H_u} + |\mu|^2$ (and not the individual parameters 
$m^2_{H_u}$ or $|\mu|^2$), just as called for in the preceding paragraph. 

In this paper, I will re-examine this possibility, with particular attention to the
previously neglected fact that there is an interplay between the hidden sector 
and the visible
sector contributions to the renormalization group (RG) running, which can lead
to quasi-fixed point relations at intermediate scales. 
Here it is particularly important to take into account restrictions 
\cite{Poland:2011ey,Poland:2015mta,Green:2012nqa}
on scaling of operators in 
superconformal field theories, which 
follow from unitarity and crossing symmetry, and which
constrain the extent to which $m^2_{H_u} + |\mu|^2$
can run. 
The emphasis is on
the little hierarchy problem that has been exacerbated during the years
of LHC searches and by the measurement of the Higgs boson mass at 125 GeV. 

\section{Review of scalar sequestering\label{sec:review}}
\setcounter{equation}{0}
\setcounter{figure}{0}
\setcounter{table}{0}
\setcounter{footnote}{1}

\begin{figure}[!tb]
\begin{center}
\includegraphics[width=0.8\linewidth,angle=0]{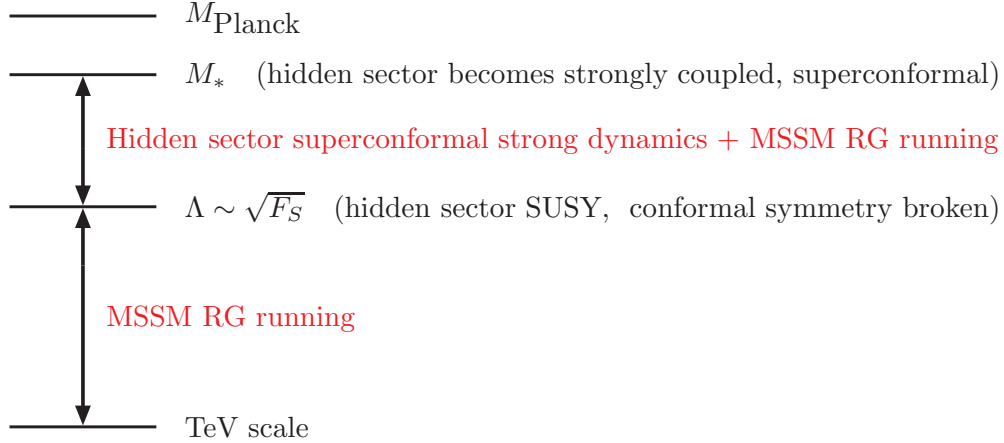}
\end{center}
\begin{minipage}[]{0.95\linewidth}
\begin{center}
\caption{\label{fig:scales}The assumed hierarchies of scales involved in the 
communication of supersymmetry breaking from the hidden sector 
(where supersymmetry is spontaneously 
broken) to the visible sector (which contains the MSSM 
particles).}
\end{center}
\end{minipage}
\end{figure}

Suppose that spontaneous supersymmetry breaking occurs in a strongly 
coupled hidden sector, and is communicated to the MSSM 
sector by a singlet chiral superfield $S$ which has a non-zero $F$-term 
VEV denoted $F_S$. The dynamics of the hidden sector, including $S$, enters
into a superconformal scaling regime at a high scale $M_*$.
The superconformal symmetry is then spontaneously broken at a lower scale to 
be denoted $\Lambda$, which is of order $\sqrt{F_S}$. 
The assumed hierarchies of scales are illustrated schematically in 
Figure \ref{fig:scales}.
The mass scale $M_*$ is ideally supposed to be much larger than $\Lambda$  (although 
this hierarchy is bounded, as pointed out in \cite{Knapen:2013zla} and discussed below).

The leading non-renormalizable terms 
that communicate supersymmetry breaking to the MSSM sector are:
\beq
{\cal L}_{\mbox{\small gaugino masses}} &=&
-\frac{c_a}{2M_*} \int d^2\theta\,  S\, {\cal W}^{a\alpha} {\cal W}^{a}_{\alpha}
 + {\rm c.c.}
\label{eq:gauginomasssource}
\\
{\cal L}_{\mbox{\small$a$\,terms}} &=&
-\frac{c^{ijk} }{6M_*} \int d^2\theta\, S \phi_i \phi_j \phi_k + {\rm c.c.}
\label{eq:atermsource}
\\
{\cal L}_{\mbox{\small$\mu$\,term}} &=&
\frac{c_\mu }{M_*} \int d^4\theta\, S^*\hspace{-0.5pt} H_u H_d + {\rm c.c.}
\label{eq:mutermsource}
\\
{\cal L}_{\mbox{\small$b$ term}} &=&
-\frac{c_b}{M_*^2} Z_{S^*\hspace{-0.5pt}S} \int d^4\theta\, S^*\hspace{-1.5pt} S H_u H_d + {\rm c.c.} 
\label{eq:btermsource}
\\
{\cal L}_{\mbox{\small$m^2$ terms}} &=&
-\frac{c_i^j}{M_*^2} Z_{S^*\hspace{-0.5pt}S} \int d^4\theta\, S^*\hspace{-1.5pt} S \phi^{* i} \phi_j
,
\label{eq:m2source}
\eeq
where the coefficients $c$ are dimensionless parameters. 
Note that there is another class of terms that could be written in the Lagrangian:
\beq
-\frac{k_i^j}{M_*} \int d^4\theta\, S \phi^{* i} \phi_j + {\rm c.c.}
\eeq
but these are redundant, as they 
can be eliminated in favor of the terms in
eqs.~(\ref{eq:btermsource}) and (\ref{eq:m2source}) 
and higher-order terms,
by a holomorphic field redefinition 
$\phi_i \rightarrow \phi_i + k_i^j S \phi_j/M_*$.

The terms in eqs.~(\ref{eq:gauginomasssource})-(\ref{eq:m2source}) 
are written in the holomorphic basis for $S$. In this basis, 
there is no hidden-sector renormalization of the
gaugino masses, $a$-terms, or the $\mu$ term, because each of 
eqs.~(\ref{eq:gauginomasssource})-(\ref{eq:mutermsource}) 
depends on either $S$ or $S^*$, but not both. In contrast, the $b$-term
and the non-holomorphic scalar squared masses are not holomorphic in $S$, and so 
are renormalized by an extra suppression factor 
\beq
Z_{S^*\hspace{-0.8pt}S} &=& (Q/Q_0)^{\Gamma}
\eeq
by the hidden-sector superconformal dynamics, where $Q$ is the renormalization scale,
$Q_0$ is a reference scale, and
\beq
\Gamma = \Delta_{S^*\hspace{-0.8pt}S} - 2 \Delta_S,
\eeq
in which 
\beq
\Delta_S = 1 + \gamma_S 
,
\eeq
where $\gamma_S$ is the anomalous dimension of $S$, and
$\Delta_{S^*\hspace{-0.8pt} S} $
is the lowest scaling dimension for a scalar operator appearing
in the operator product expansion of $S^*\hspace{-0.8pt}$ and $S$.
In this holomorphic basis, $S$ has
a non-canonical kinetic term:
\beq
{\cal L}_{\mbox{\small kinetic}} &=&
Z_S \int d^4\theta\, S^* S,
\eeq
where the hidden-sector wavefunction renormalization factor is
\beq
Z_S = (Q/Q_0)^{-2 \gamma_{\raisebox{-2pt}{$\scriptscriptstyle S$}}}.
\eeq
Now one can go to a canonical basis for $S$, by making the redefinition
\beq
S \rightarrow Z_S^{-1/2} S.
\eeq
In the canonical basis for $S$, the MSSM gaugino masses
$\fixwidetilde M_a$, scalar cubic couplings $\fixwidetilde a^{ijk}$, 
$\fixwidetilde\mu$ term, the holomorphic supersymmetry-breaking Higgs squared mass term 
$\fixwidetilde b$, and 
non-holomorphic supersymmetry breaking scalar squared masses can be evaluated 
as:\footnote{The reason for the tildes in the names of these dimensionful 
parameters is to distinguish them from their counterparts after 
a further redefinition to be made shortly.}
\beq
\fixwidetilde M_a &=& c_a Z_S^{-1/2} F_S/M_*,
\\
\fixwidetilde a^{ijk} &=& c^{ijk} Z_S^{-1/2} F_S/M_* ,
\\
\fixwidetilde \mu &=& c_\mu Z_S^{-1/2} F_S^*/M_*,
\\
\fixwidetilde b  &=& c_b Z_{S^*\hspace{-0.5pt}S} Z_S^{-1} |F_S|^2/M_*^2,
\\
(\fixwidetilde m^2)_i^j &=& c_i^j Z_{S^*\hspace{-0.5pt}S} Z_S^{-1} |F_S|^2/M_*^2.
\eeq
The presence of the extra hidden-sector renormalization factor $Z_{S^*S}$ for the
MSSM parameters of mass dimension two implies that they have a modified running for the
range $\Lambda < Q < M_*$, throughout which the hidden sector is assumed to be 
nearly superconformal. 

Now if we use 
the generic notations $\fixwidetilde M_A$ and $ \fixwidetilde m^2_i$ for 
parameters of mass dimensions one and two respectively, and take into account the
visible sector renormalization in a canonical basis for the MSSM fields, then
the renormalization group running for $Q>\Lambda$ is: 
\beq
\frac{d}{dt} \fixwidetilde M_{\!A}
&=& \gamma_{\raisebox{-2pt}{\scriptsize$S$}} \fixwidetilde M_{\!A} + 
\beta^{\raisebox{2pt}\MSSM}_{\raisebox{-2pt}{$\scriptstyle\fixwidetilde M_{\!A}$}}
,
\\
\frac{d}{dt} \fixwidetilde m^2_i
&=&  
(\Gamma + 2 \gamma_S) \fixwidetilde m^2_i + 
\beta^{\raisebox{2pt}\MSSM}_{\raisebox{-2pt}{$\scriptstyle\fixwidetilde m^2_i$}}
.
\eeq
where
$\beta^{\raisebox{2pt}\MSSM}_{\raisebox{-2pt}{$\scriptstyle\fixwidetilde M_{\!A}$}}$ 
and $\beta^{\raisebox{2pt}\MSSM}_{\raisebox{-2pt}{$\scriptstyle\fixwidetilde m^2_i$}}$
are the usual beta functions obtained without including hidden sector effects, and
\beq
t \equiv \ln(Q/Q_0).
\eeq

To simplify the renormalization group running in practice, it is convenient to make a redefinition to undo
the effect of going to the canonical basis for $S$, but remain in the canonical basis for
the MSSM fields, by now defining, for $Q \geq \Lambda$:
\beq
M_{\!A} &=& (\Lambda/Q)^{\gamma_S}\, \fixwidetilde M_{\!A} , 
\\
m^2_i &=& (\Lambda/Q)^{2 \gamma_S}\, \fixwidetilde  m^2_i,
\eeq
which then run according to:
\beq
\frac{d}{dt} M_A
&=& \beta^{\raisebox{1pt}\MSSM}_{M_A} ,
\label{eq:betaMhybrid}
\\
\frac{d}{dt} m^2_i
&=& \Gamma m^2_i + \beta^{\raisebox{1pt}\MSSM}_{m^2_i}
,
\label{eq:betamhybrid}
\eeq
where it is easy to check using dimensional analysis 
that 
$\beta^{\raisebox{1pt}\MSSM}_{M_A}$ and $\beta^{\raisebox{1pt}\MSSM}_{m^2_i}$
are obtained from 
$\beta^{\raisebox{2pt}\MSSM}_{\raisebox{-2pt}{$\scriptstyle\fixwidetilde M_{\!A}$}}$ 
and $\beta^{\raisebox{2pt}\MSSM}_{\raisebox{-2pt}{$\scriptstyle\fixwidetilde m^2_i$}}$ 
by simply substituting $\fixwidetilde M_{\!A}\rightarrow M_A $ and
$\fixwidetilde m_i^2 \rightarrow m^2_i$.
Note that $\gamma_S$ has thus been eliminated from the running.
For simplicity, $\Gamma$ is taken here to be a positive constant for 
$Q>\Lambda$, corresponding to an
idealized  exactly superconformal theory in the hidden sector, while $\Gamma=0$ for
$Q<\Lambda$ where the superconformal symmetry of the hidden sector is broken.
At the scale $Q=\Lambda$, the parameters are assumed to simply match, although
in a more complete realistic model 
they are likely governed by more complicated threshold corrections, and $\Gamma$
will not be exactly constant.

\baselineskip=17pt

A crucial subtlety is that the $\mu$ term hidden sector renormalization feeds 
\cite{Roy:2007nz,Murayama:2007ge,Perez:2008ng}
into that
of the non-holomorphic Higgs scalar squared masses, 
so that the combinations that are subject to the hidden-sector scaling are actually
\cite{Murayama:2007ge,Perez:2008ng} the full non-holomorphic scalar squared mass combinations:
\beq 
\fixwidehat m^2_{H_u} &\equiv& m^2_{H_u} + |\mu|^2,
\\
\fixwidehat m^2_{H_d} &\equiv& m^2_{H_d} + |\mu|^2,
\eeq
rather than the supersymmetry-breaking 
parameters $m^2_{H_u}$ and $m^2_{H_d}$. 
Equations~(\ref{eq:betaMhybrid}) and (\ref{eq:betamhybrid})
therefore apply to the MSSM parameters:
\beq
M_A &=& \mbox{gaugino masses, $a$ terms, and the $\mu$ term,}
\label{eq:listMa}
\\
m_i^2 &=& \mbox{squark and slepton squared masses, 
$\fixwidehat m^2_{H_u}$, $\fixwidehat m^2_{H_d}$, and $b$}.
\label{eq:listmi2}
\eeq
In the remainder of this paper, I will stick to the scheme in which
eqs.~(\ref{eq:betaMhybrid}) and (\ref{eq:betamhybrid}) hold, 
with boundary conditions for the input parameters of 
eqs.~(\ref{eq:listMa}) and (\ref{eq:listmi2}) 
to be specified at the scale
$Q=M_*$. [At that scale, one has the equivalences
$\fixwidetilde M_{\!A} = (M_*/\Lambda)^{\gamma_S} M_A$ and
$\fixwidetilde m^2_i = (M_*/\Lambda)^{2 \gamma_S} m^2_i$, while at the matching scale
$Q=\Lambda$, $\fixwidetilde M_{\!A} = M_A$ and
$\fixwidetilde m^2_i = m^2_i$.]

In the numerical results below, I will use the 2-loop MSSM beta functions found in 
refs.~\cite{Martin:1993zk,Yamada:1994id,Jack:1994kd,Jack:1994rk}.
The $\fixwidehat m^2_{H_u}$ and $\fixwidehat m^2_{H_d}$ MSSM beta functions
are obtained straightforwardly from these, for example:
\beq
\beta^{\raisebox{2pt}\MSSM}_{\fixwidehat m^2_{H_u}}
&=& 
\frac{1}{16 \pi^2} \Bigl [6 y_t^2 (\fixwidehat m^2_{H_u} + m^2_{Q_3}
+ m^2_{{u}_3}) + 
(6 y_b^2 + 2 y_\tau^2) \mu^2 + 6 a_t^2
\nonumber \\ &&
- 6 g_2^2 (M_2^2 + \mu^2)
- \frac{6}{5} g^2_1 (M_1^2 + \mu^2 - T/2)
\Bigr ] + \ldots 
,
\label{eq:betam2Huhat}
\\
\beta^{\raisebox{2pt}\MSSM}_{\fixwidehat m^2_{H_d}}
&=& 
\frac{1}{16 \pi^2} \Bigl [6 y_b^2 (\fixwidehat m^2_{H_d} + m^2_{Q_3}
+ m^2_{{d}_3}) +
2 y_\tau^2  (\fixwidehat m^2_{H_d} + m^2_{L_3}
+ m^2_{\overline{e}_3})
\nonumber \\ &&
+
6 y_t^2 \mu^2 + 6 a_b^2 + 2 a_\tau^2 
- 6 g^2_2 (M_2^2 + \mu^2)
- \frac{6}{5} g^{2}_1 (M_1^2 + \mu^2 + T/2)
\Bigr ] + \ldots
,
\phantom{xxx}
\label{eq:betam2Hdhat}
\eeq
where the ellipses represent the contributions beyond 1-loop order, 
and $g_1$ and $g_2$ are the electroweak gauge couplings in a 
grand unified theory (GUT) normalization,
and 
\beq
T =\fixwidehat m^2_{H_u} - \fixwidehat m^2_{H_d} + 
\sum_{i=1}^3 [m^2_{Q_i} - m^2_{L_i} - 2 m^2_{u_i} + m^2_{d_i} + m^2_{e_i}]
.
\eeq

\section{Quasi-fixed points from interplay of hidden and visible renormalization\label{sec:quasi}}
\setcounter{equation}{0}
\setcounter{figure}{0}
\setcounter{table}{0}
\setcounter{footnote}{1}

In earlier work, it has often been assumed that $\Gamma$ is large and positive, 
so that the MSSM contributions to the
running of the dimension two parameters are relatively negligible for $Q > \Lambda$. 
In the idealized limit of large $\Gamma$, 
there is power-law running resulting in 
a relative suppression $(\Lambda/M_*)^\Gamma$ for the dimension two terms, 
compared to the squares of dimension one terms, 
at the scale $Q = \Lambda$. In that limit, one naively can impose boundary conditions 
\beq
m_i^2 \approx 0
\label{eq:bogusbc}
\eeq
at $Q = \Lambda$, provided that there is a significant hierarchy 
$\Lambda/M_*$. 

However, constraints on superconformal field theories 
have shown \cite{Poland:2011ey,Poland:2015mta} that while $\Gamma$ indeed 
might be positive, it cannot be 
too large, with stronger bounds for smaller $\gamma_S$. These papers 
have also provided some circumstantial evidence for the existence of
a minimal superconformal theory, which may (based on 
extrapolation of established constraints) have
\beq
\gamma_S &\approx& 3/7 ,
\\
\Gamma &\lsim& 0.3,
\eeq 
although there is so far no specific 
identification of this theory or guarantee of its existence.
The constraints found in \cite{Poland:2011ey,Poland:2015mta,Green:2012nqa} 
also imply that any 
significantly smaller $\gamma_S$ would necessarily have a much smaller $\Gamma$.

From these constraints, in any given model there is a limit \cite{Knapen:2013zla}
on the range of scales at which the hidden sector can remain in the
superconformal regime, based on existing LHC bounds on the superpartner masses. 
Taking $\Lambda = \sqrt{F_S}$ and $c_3$ of order unity, and requiring that
at the scale $Q=\Lambda$ the gluino mass 
$M_3 \sim c_3 (F_S/M_*) (\Lambda/M_*)^{\gamma_S}$
exceeds 1000 GeV, this constraint amounts to roughly:
\beq
\Lambda &\gsim& \left [(\mbox{1000 GeV}) 
M_*^{1 + \gamma_S}  \right ]^{1/(2 + \gamma_S)}
\eeq
Now taking $\gamma_S \gsim 3/7$, and identifying $M_*$ with the scale
$M_{\rm GUT} = 2.5 \times 10^{16}$ GeV at which the gauge couplings appear to unify,
one finds that $\Lambda \gsim 8 \times 10^{10}$ GeV. 

In the following, I will therefore optimistically take 
$M_* = M_{\rm GUT}$ and $\Lambda = 10^{11}$ GeV and $\Gamma= 0.3$ for 
numerical examples.\footnote{Anomaly-mediation \cite{AMSB} contributions
of order $F_S/16 \pi^2 M_{\rm Planck}$ to gaugino masses and
$F_S^2/(16 \pi^2 M_{\rm Planck})^2$ to scalar squared masses are neglected here,
but could be significant for larger choices of $M_*$.}
For this, or smaller, values of $\Gamma$, 
a more accurate treatment than eq.~(\ref{eq:bogusbc}) 
is that
the dimension two MSSM parameters are drawn towards a quasi-fixed point trajectory 
solution at which the 
MSSM contributions to the beta functions balance with the hidden sector contributions, so
that the right side of eq.~(\ref{eq:betamhybrid}) approximately vanishes. 
The quasi-fixed point trajectories for the dimension two parameters 
therefore can be roughly approximated as the solutions of the algebraic equations
\beq
m^2_{i,\,{\mbox{\small quasi-fixed}}} &\approx&
- \beta^{\raisebox{1pt}\MSSM}_{m^2_i}/\Gamma
,
\label{eq:qfpgeneral}
\eeq
where the dimension two parameters appearing in the MSSM beta functions
on the right-hand side are self-consistently set
equal to their quasi-fixed point values. 
However, note that these quasi-fixed points 
are moving targets, which in practice means that the right-hand side of 
eq.~(\ref{eq:qfpgeneral})
should be evaluated at a scale slightly larger than the left-hand side. 
Equation~(\ref{eq:qfpgeneral}) does provide useful approximate relations discussed in the next few paragraphs, but rather than attempting a more detailed and precise 
analytical expression for the
quasi-fixed point trajectories, in the examples of the next section the running 
will be evaluated numerically.

Fortunately, the MSSM beta functions for squark and slepton squared 
masses come mostly from gaugino masses, and are negative. This means 
that the squared masses will tend to approach positive quasi-fixed point
values at the matching scale $Q = \Lambda$. 
For example, the rough estimate for the quasi-fixed point of the running
right-handed selectron mass, in terms of the running bino mass 
parameter $M_1$
and the $U(1)_Y$ gauge coupling in a GUT normalization $g_1$, is:
\beq
m_{\tilde e_R, {\mbox{\small quasi-fixed}}} 
&\approx& 
\sqrt{\frac{3}{10}} \frac{g_1 M_1}{\pi \sqrt{\Gamma}} 
=
0.18 
\left (\frac{g_1}{0.57} \right ) \left(\frac{0.3}{\Gamma} \right )^{1/2} M_1
,
\label{eq:qfpsleptons}
\eeq
rather than 0, and for a typical squark mass, including only the effects of the running 
gluino mass $M_3$:
\beq
m_{\tilde q, {\mbox{\small quasi-fixed}}} 
&\approx& 
\sqrt{\frac{2}{3}} \frac{g_3 M_3}{\pi \sqrt{\Gamma}} 
=
0.365
\left (\frac{g_3}{0.77} \right ) \left(\frac{0.3}{\Gamma} \right )^{1/2} M_3
.
\label{eq:qfpsquarks}
\eeq
These expressions have been normalized to typical values for the 
MSSM running gauge
couplings at an intermediate scale $\Lambda = 10^{11}$ GeV.

However, as noted above, 
the quasi-fixed points are moving targets, because the gaugino masses 
(and other contributions to the MSSM beta functions) are running with the scale $Q$.
Another important practical effect is that in realistic models
the quasi-fixed point trajectories are not actually reached 
with the finite running available
from $M_*$ down to $\Lambda$, 
so the above estimates are not quite realized. As we will see, at $Q=\Lambda$
the running masses of the squarks and 
especially the sleptons are often considerably higher
than the estimates of eqs.~(\ref{eq:qfpsleptons}) 
and (\ref{eq:qfpsquarks}) would indicate. There is also a significant effect due to
the subsequent running
from the scale $\Lambda$ down to the TeV scale. The effect of running below 
$\Lambda$ is actually the dominant effect for the physical squark masses (because
$M_3$ and $g_3$ are growing in the infrared), 
but it is relatively much smaller for sleptons (because $M_1$ and $g_1$ are 
shrinking in the infrared). Therefore, the
influence of the quasi-fixed point behavior given by eq.~(\ref{eq:qfpgeneral}) 
turns out to be crucial for understanding how sleptons can be 
heavy enough to avoid discovery at the LHC, and also
how the LSP need not be a charged slepton.

The quasi-fixed point behavior of the Higgs squared masses is of even greater importance.
From eqs.~(\ref{eq:betam2Huhat}) and (\ref{eq:qfpgeneral}), 
one obtains an estimate for the running 
quasi-fixed point trajectory, for $Q$ much less than $M_*$ 
but not smaller than  $\Lambda$:
\beq
\fixwidehat m^2_{H_u, {\mbox{\small quasi-fixed}}} 
&\approx& \frac{3}{8 \pi^2 \Gamma + 3 y_t^2}
\Bigl [
g_2^2 (M_2^2 + \mu^2) + \frac{g_1^2}{5} (M_1^2 + \mu^2 - T/2) 
- a_t^2 
\nonumber \\ && 
- y_t^2 (m^2_{Q_3} + m^2_{u_3})
- \mu^2 (y_b^2 + y_\tau^2/3)
\Bigr ] .
\label{eq:qfpm2Huhat}
\eeq
Realistic electroweak symmetry breaking requires that 
$\fixwidehat m^2_{H_u}$ must be small in magnitude
near the TeV scale; 
this is the essence of the supersymmetric little hierarchy problem. Because
eq.~(\ref{eq:qfpm2Huhat}) has significant 
contributions of both signs, and is suppressed by $8 \pi^2 \Gamma$, 
one can adjust it to the appropriate value, 
even if $a_t$, $\mu$, $M_2$, $M_1$, 
$m_{Q_3}$ and $m_{u_3}$ are much larger in magnitude than a TeV. 
As a guide to finding models with correct electroweak symmetry breaking, 
note that to decrease the low-energy prediction for $\fixwidehat m^2_{H_u}$, 
one can increase $|a_t|$, $m^2_{Q_3}$, or $m^2_{u_3}$,
or decrease $|\mu|$, $|M_2|$, or $|M_1|$. 
However, the subsequent running from $Q=\Lambda$ down to the TeV scale is 
also quite significant, so that the actual value of
$\fixwidehat m^2_{H_u}$ that is needed at $Q=\Lambda$ 
is not an extremely small value. The requirement of
correct electroweak symmetry breaking can easily be obtained by adjusting 
the input parameters in a predictive way, but the level of tuning
required, while arguably reduced as illustrated in examples below, 
cannot be said to be eliminated. 

The other dimensionful quantities appearing in the Higgs potential, 
$\fixwidehat m^2_{H_d}$ and $b$, are also strongly influenced 
to flow towards quasi-fixed point trajectories
in the infrared. This implies that to some approximation, the 
information about the initial conditions at $Q=M_*$ is washed out, and roughly speaking 
the Higgs potential
parameters are predicted in terms of the dimension one parameters of the theory. 
Despite the fact that this washing out is incomplete due to the quasi-fixed points not
quite being reached, there are some robust trends that remain in the form of 
constraints and correlations
between the soft supersymmetry-breaking parameters of the theory at the TeV scale. 
These will be explored numerically for some sample slices in parameter space in the next
section.

\section{Numerical examples}
\setcounter{equation}{0}
\setcounter{figure}{0}
\setcounter{table}{0}
\setcounter{footnote}{1}

In this section, I provide some examples that illustrate the importance of the
quasi-fixed point behavior in the presence of scalar sequestering dynamics.  
The low-energy results are only weakly dependent on the high-scale boundary condition
values of the dimension two parameters. 
Therefore, for simplicity and to keep the dimensionality of parameter space small, 
I will take all of the squark and slepton squared masses and 
$\fixwidehat m_{H_u}^2$ and $\fixwidehat m_{H_d}^2$ 
to be equal to a common value $m_0^2$ at $Q=M_*$.
Also for simplicity,\footnote{This is a quite non-trivial assumption
from the point of view of the supersymmetric flavor problem, since completely general
scalar cubic interactions could be dangerous. 
However, imposing a flavor symmetry on these terms
is technically natural, and the contributions to scalar squared masses 
mediated by RG running from the gaugino masses are flavor-blind, and could dominate
the sfermion mixings.}
I assume that the scalar trilinear couplings
are governed by a universality condition, with $a_t = A_0 y_t$ and $a_b = A_0 y_b$ and
$a_\tau = A_0 y_\tau$ at $Q=M_*$; with this assumption, the particular values 
of $a_b$ and $a_\tau$ are of much less importance than that of $a_t$. 
The dimensionful input parameters are therefore
\beq
M_1,\, M_2,\, M_3,\, A_0,\, \mu,\, m_0^2,\, b,
\eeq
specified at the renormalization scale $Q = M_*$. 
The RG scale at which superconformal
running begins in the hidden sector is $M_* =  M_{\rm GUT} = 2.5 \times 10^{16}$ GeV,
and the superconformal running regime ends at RG scale $\Lambda = 10^{11}$ GeV. 
The parameters of the theory are run using  
eqs.~(\ref{eq:betaMhybrid}) and (\ref{eq:betamhybrid}) 
with $\Gamma = 0.3$ from $Q=M_*$ down to $Q=\Lambda$, and then with 
$\Gamma = 0$ from $Q=\Lambda$ to the electroweak scale.

As a constraint on the parameter space, the requirement of correct electroweak 
symmetry breaking with $m_Z = 91$ GeV and a fixed value of $\tan\beta$ are imposed. 
This allows for two parameters to be solved for, 
which I take to be $A_0$ and $\mu$. (It is better to avoid choosing $m^2_0$ or $b$ 
as a parameter to be solved for, due to the much weaker dependence of 
the Higgs potential on their high-scale values, 
which follows from the quasi-fixed point behavior.)
In practice, the parameters $A_0$ and $\mu$ are solved for 
by iterating to convergence, starting from an arbitrary initial guess.
The lightest Higgs boson mass $M_h$ is obtained using the leading 
3-loop calculation given in 
\cite{Martin:2007pg}, augmented by 1-loop electroweak corrections. 
In recognition of the theoretical and parametric uncertainties
\cite{Mhuncertainties}
in the $M_h$ calculation, this predicted value is
required to be in the range
$\mbox{123 GeV} < M_h < \mbox{127 GeV}$. The $M_h$ constraint
has the greatest impact on the allowed values of $M_3$ (which feeds into 
the magnitudes of the top-squark masses) and $A_0$ (which controls the top-squark mixing).
Search limits on direct superpartner production 
at the LHC turn out to not constrain the models given below, because 
the Higgs mass constraint indirectly
requires the gluino and squarks to be heavy anyway.

\subsection{Pessimistic case: unified gaugino mass boundary conditions
\label{subsec:unified}}

In this subsection I consider models that have unified gaugino mass
parameters $M_1 = M_2 = M_3 = m_{1/2}$ at $Q=M_*$, and $\tan\beta=15$ fixed. 
The electroweak symmetry breaking constraint then turns out to require that 
$A_0$ is positive (see Figure \ref{fig:AmuXt} below), which in turn 
implies that at the electroweak scale 
$a_t$ is negative but not very large in magnitude, so that 
top-squark mixing is moderate. Because of this, to obtain 
$M_h$ in agreement with experiment 
requires a rather large $m_{1/2}$, between about 2.7 TeV 
and 8.3 TeV to obtain an estimated
123 GeV $<M_h<$ 127 GeV.

\begin{figure}[!tb]
\begin{minipage}[]{0.49\linewidth}
\includegraphics[width=\linewidth,angle=0]{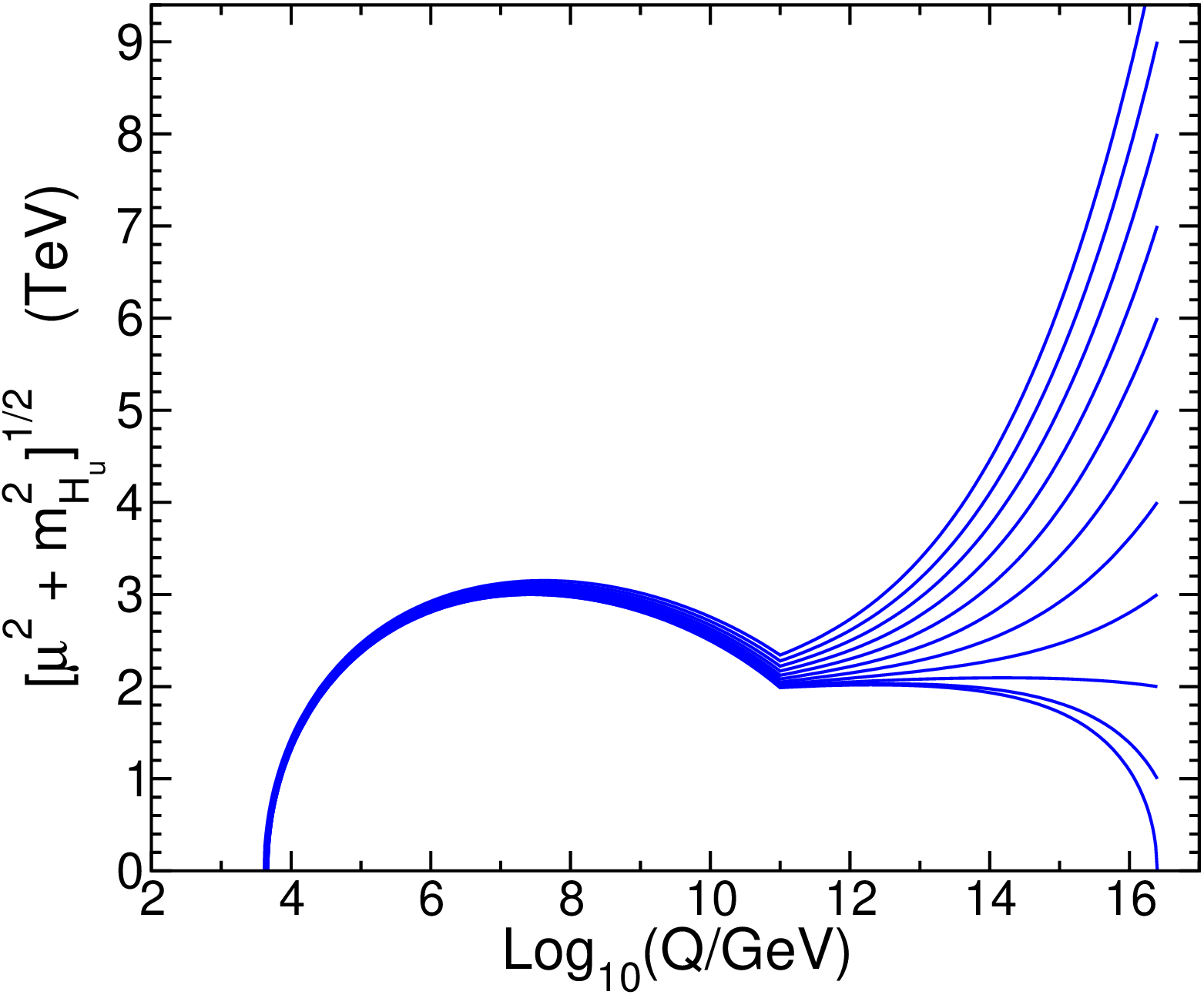}
\end{minipage}
\begin{minipage}[]{0.49\linewidth}
\begin{flushright}
\includegraphics[width=\linewidth,angle=0]{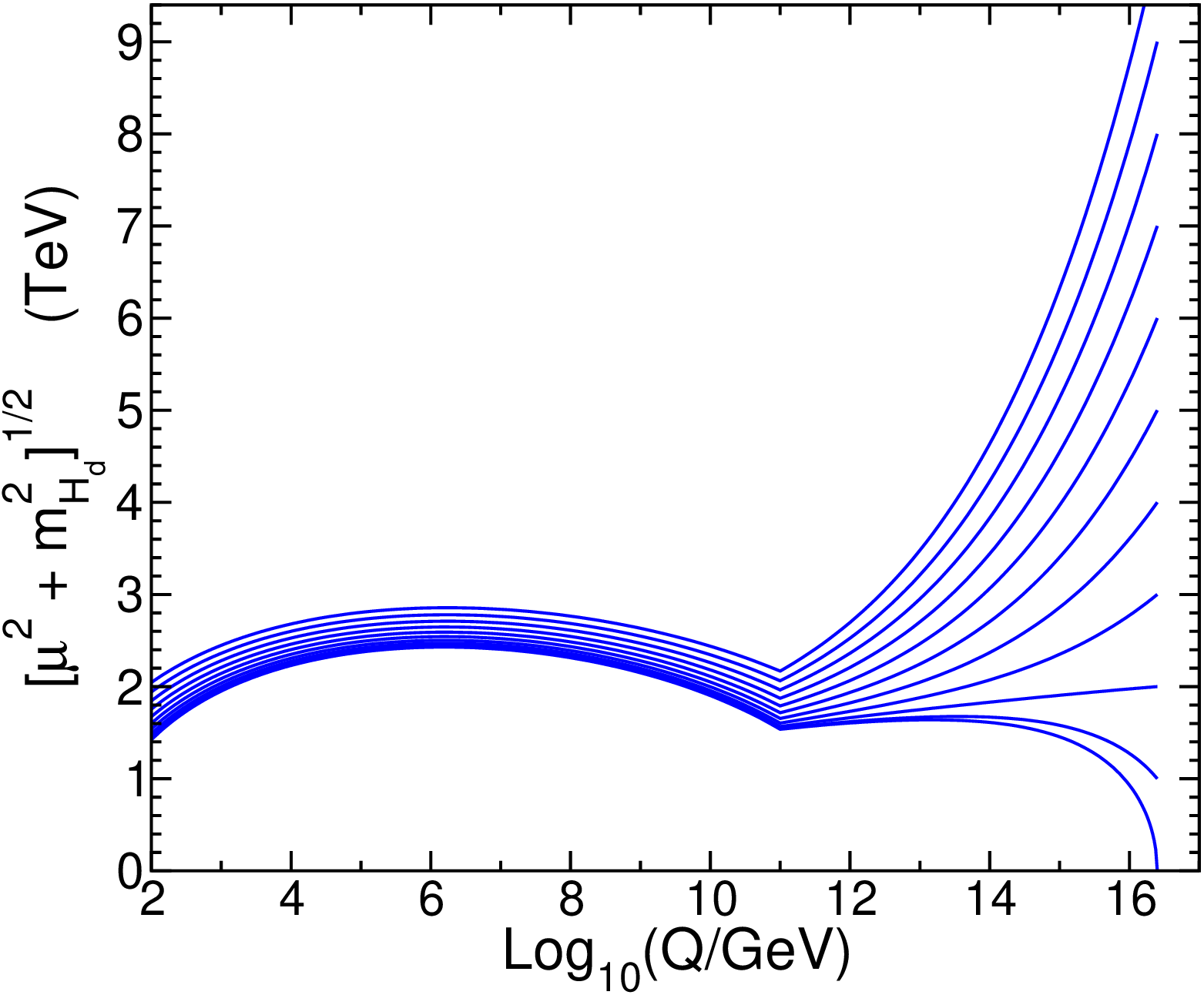}
\end{flushright}
\end{minipage}
\begin{minipage}[]{0.49\linewidth}
\includegraphics[width=\linewidth,angle=0]{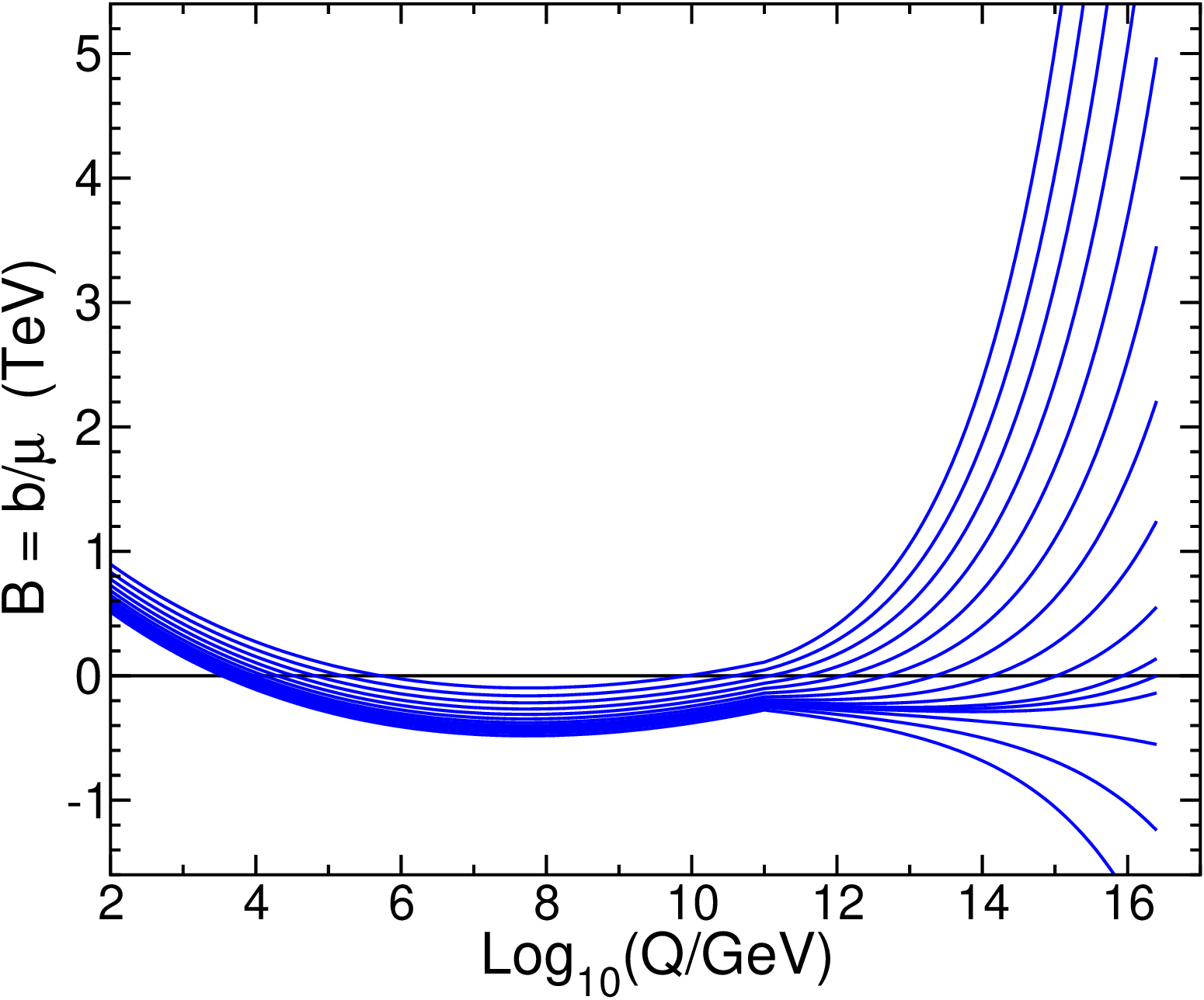}
\end{minipage}
\hspace{0.04\linewidth}
\begin{minipage}[]{0.45\linewidth}
\caption{\label{fig:runQ_U_Higgs}
Renormalization group running of the Higgs mass parameters 
$(m^2_{H_u} + \mu^2)^{1/2}$, 
$(m^2_{H_d} + \mu^2)^{1/2}$, and 
$B = b/\mu$,  as a function of the RG scale $Q$,
for a model with $M_{1} = M_{2} = M_{3} = 4500$ GeV at $M_{\rm GUT}$, 
and $\Gamma = 0.3$ and $\Lambda = 10^{11}$ GeV. Each line is the RG 
trajectory for a 
different boundary condition of 
the common scalar squared mass $m_0^2 = b$ at the GUT scale, showing the 
approach to the quasi-fixed point behavior at low RG scales.}
\end{minipage}
%

\vspace{-0.3cm}

\begin{center}
\noindent\rule{1.01\linewidth}{0.8pt}
\end{center}

\begin{minipage}[]{0.49\linewidth}
\includegraphics[width=\linewidth,angle=0]{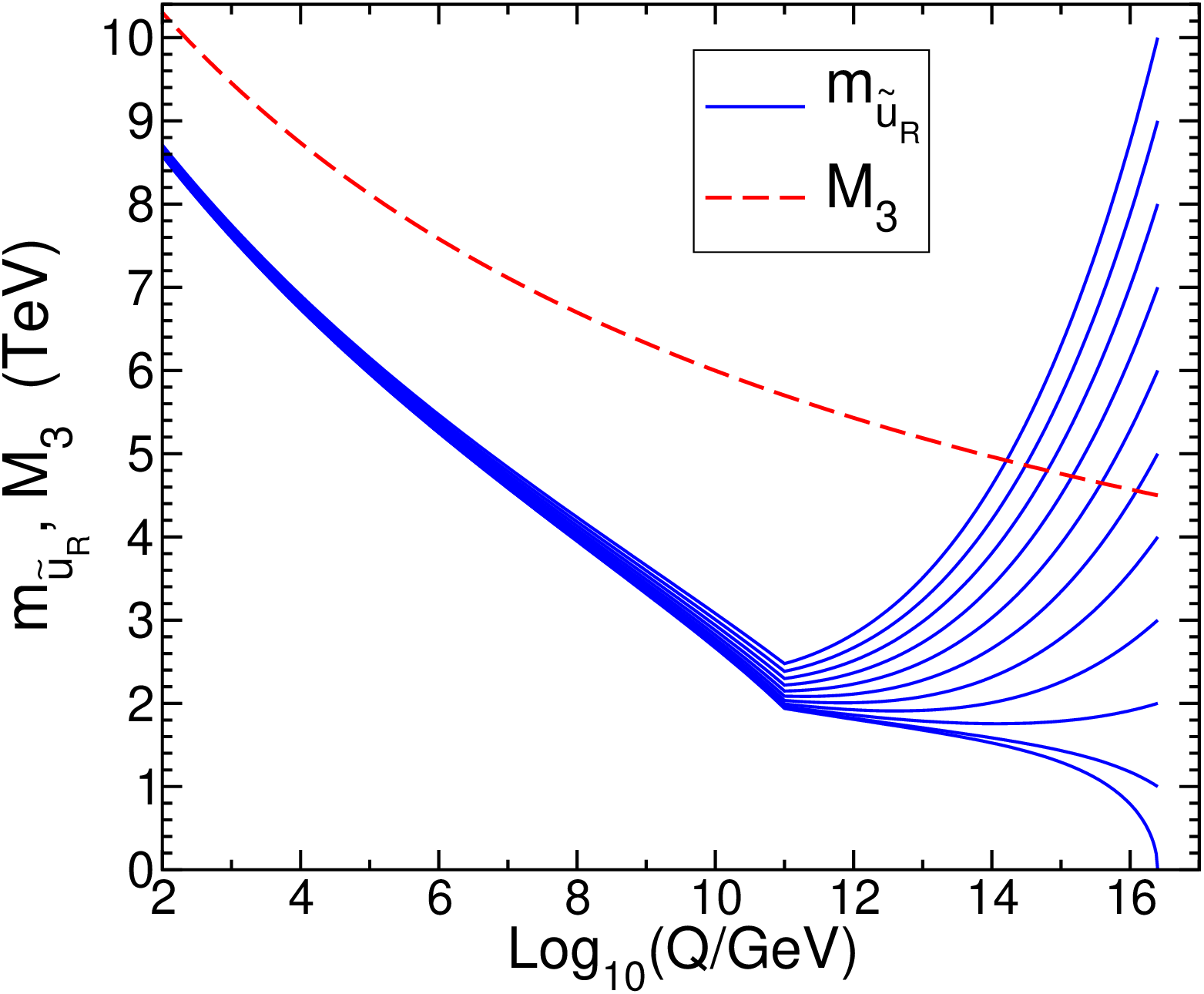}
\end{minipage}
\begin{minipage}[]{0.49\linewidth}
\begin{flushright}
\includegraphics[width=\linewidth,angle=0]{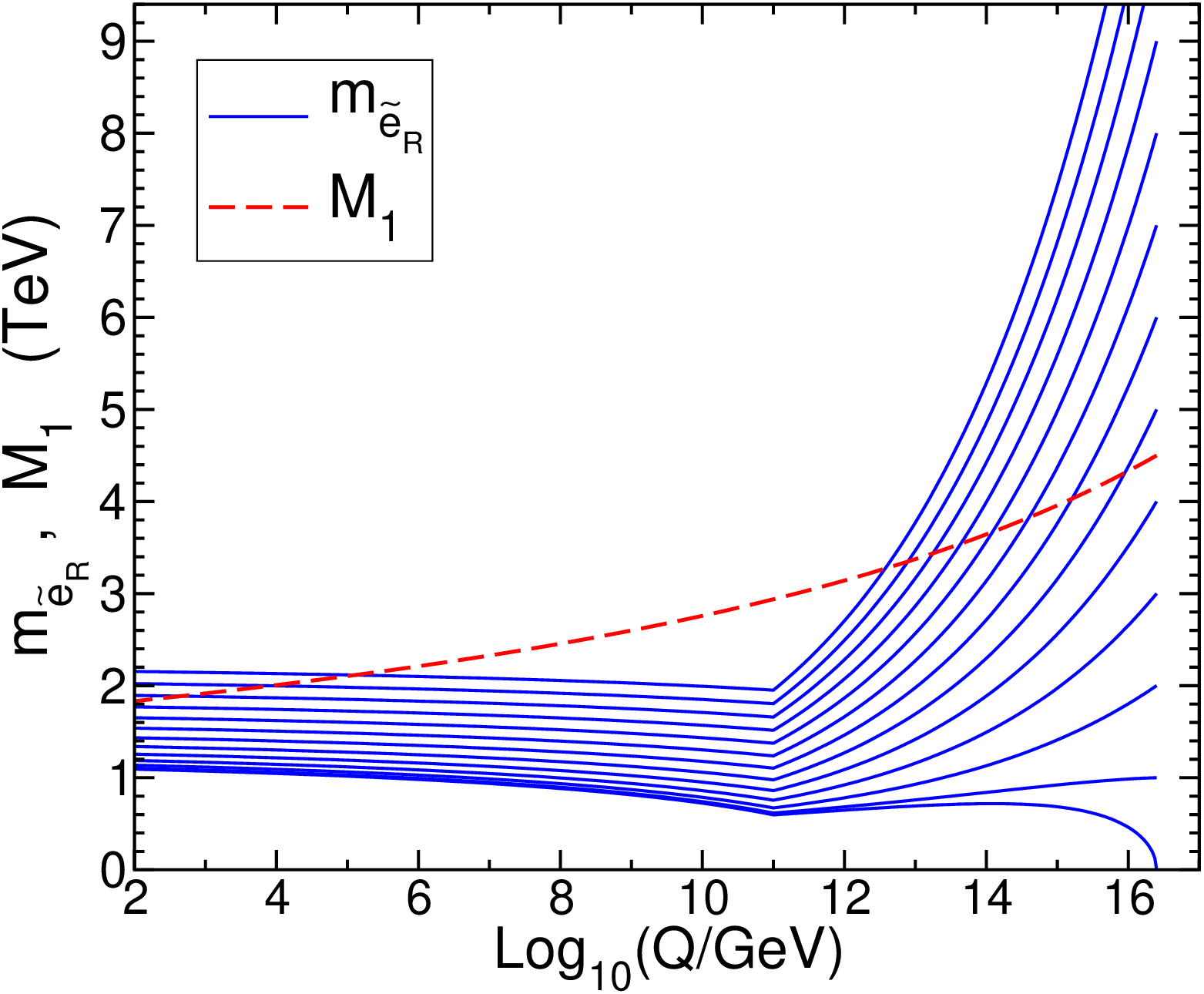}
\end{flushright}
\end{minipage}
\caption{\label{fig:runQ_U_sfermions} 
RG running of the squark and slepton mass parameters 
$m_{\tilde u_R}$ (left panel) and $m_{\tilde e_R}$ (right panel) 
for a model with
$M_{1} = M_{2} = M_{3} = 4500$ GeV at $M_{\rm GUT}$, and $\Gamma = 0.3$ 
and $\Lambda = 10^{11}$ GeV. Each line is the RG trajectory for a 
different boundary condition of the common scalar mass at the GUT 
scale, showing the approach to the quasi-fixed point 
behavior. The running gluino and bino masses $M_3$ and $M_1$ are 
also shown for comparison.}
\end{figure}

The quasi-fixed point behavior of the RG running with $Q$ is shown for the choice
$m_{1/2} = 4500$ GeV, for various boundary condition values of the
remaining independent input parameter $m_0^2 = b$, in Figures \ref{fig:runQ_U_Higgs}
and \ref{fig:runQ_U_sfermions}. 
The trajectories for $\fixwidehat m_{H_u}^2$ converge to quasi-fixed 
point values near (2 TeV)$^2$ in this model, in good agreement 
with eq.~(\ref{eq:qfpm2Huhat}). 
The fact that this is not 
much smaller illustrates the importance of including the visible sector 
renormalization effects for $Q>\Lambda$
together with the hidden sector scaling. Even for very 
large $m_0^2$, shown up to (10 TeV)$^2$ in the figure, the value of 
$\fixwidehat m_{H_u}^2$ at $Q = \Lambda$ is only slightly higher. The running 
of $\fixwidehat m_{H_u}^2$ down to the TeV scale is quite substantial, 
but Figure \ref{fig:runQ_U_Higgs}a shows that there is an additional 
focusing effect for $Q<\Lambda$ that helps to make the low-energy trajectory 
rather insensitive to the high-scale value of $m_0$. However, the 
model cannot be viewed as free of fine-tuning, because the 
values of $A_0$ and $\mu$ have to be chosen rather precisely in this 
case to ensure that $\fixwidehat m_{H_u}^2$ runs close to 0 at the 
appropriate RG scale of a few TeV in order to obtain $m_Z = 91$ GeV. This
can be appreciated by noting the large magnitude of the slope of the running of 
$\fixwidehat m_{H_u}^2$ for $Q$ below the 10 TeV scale. 

The other two panels of Figure \ref{fig:runQ_U_Higgs} show the 
quasi-fixed point behavior for the other two dimensionful Lagrangian 
parameters appearing in the tree-level Higgs potential, $\fixwidehat m_{H_d}^2$ and 
$B = b/\mu$. Note that $\fixwidehat m_{H_d}^2$ also flows to values of 
order (2 TeV)$^2$ at the intermediate scale, but is not as strongly 
modified by the running below this. The running of $B$ flows towards 
small negative values at the intermediate scale, and it runs positive 
near the TeV scale but does not exceed a few hundred GeV in magnitude. 
In both of these cases, the focusing (both from the quasi-fixed point 
nature of the running above $\Lambda$ and the subsequent running below 
$\Lambda$) is not as pronounced as for $\fixwidehat m_{H_u}^2$.

The first panel of Figure \ref{fig:runQ_U_sfermions} shows the quasi-fixed point 
influence on the running of a typical squark mass, in this case $\fixwidetilde u_R$. 
For comparison, the
running gluino mass parameter $M_3$, which mostly drives the squark masses, 
is also shown. Again, the
quasi-fixed point value of the squark masses at $Q=\Lambda$
is of order 2 TeV [somewhat larger than
the rough prediction of eq.~(\ref{eq:qfpsquarks})], but the effect of pure MSSM
running for
$Q<\Lambda$ dominates over this, resulting in squark masses at the weak scale that are of
order $0.8 M_3$, even if the initial value $m_0$ is much larger or smaller. This is a robust prediction of the framework, and it depends only weakly on $\Gamma$ or $\Lambda$,
provided only that the latter is not too small. This is qualitatively similar to models
with no-scale \cite{Ellis:1984bm,Ellis:1983sf,Lahanas:1986uc}
or gaugino-mediated \cite{Kaplan:1999ac,Chacko:1999mi,Schmaltz:2000gy}
boundary conditions.

The running of the right-handed selectron mass is shown in the right panel of 
Figure \ref{fig:runQ_U_sfermions}. Here it is apparent 
that the approach to the quasi-fixed point
trajectory is much slower, resulting in a much larger spread of possible values for 
$m_{\tilde e_R}$ at $\Lambda$ [which generally exceed the rough prediction
of eq.~(\ref{eq:qfpsleptons})]
and at the TeV scale. For comparison, also shown is 
the bino mass parameter $M_1$, which is mainly responsible for driving it. 
There 
is a competition for the role of the LSP between the lightest charged 
slepton and a bino-like neutralino. If $m_0 \lsim 2.5 m_{1/2}$, 
then the LSP will be a charged slepton. 
To avoid cosmological problems from a charged stable LSP, one can 
invoke $R$-parity violation to allow the LSP to decay. 
Conversely, if $m_0 \gsim 2.5 m_{1/2}$ at $Q=M_*$, then the LSP will be a neutralino, and
could in principle be the dark matter, if $R$-parity is conserved. 
Obtaining a correct thermal relic abundance from the 
early universe may require some fine adjustment of the masses, 
to enable the stau co-annihilation mechanism for example. 
The dividing line between 
these two cases is rather robust and generally given by 
$m_{{\tilde \ell}_R} \approx 2.5 M_1$ at $Q=M_*$
even in models without gaugino mass unification,
because to a good approximation only $M_1$ enters into the 
quasi-fixed point attraction and subsequent running below $\Lambda$ for the 
right-handed slepton masses.
 
\begin{figure}[!tb]
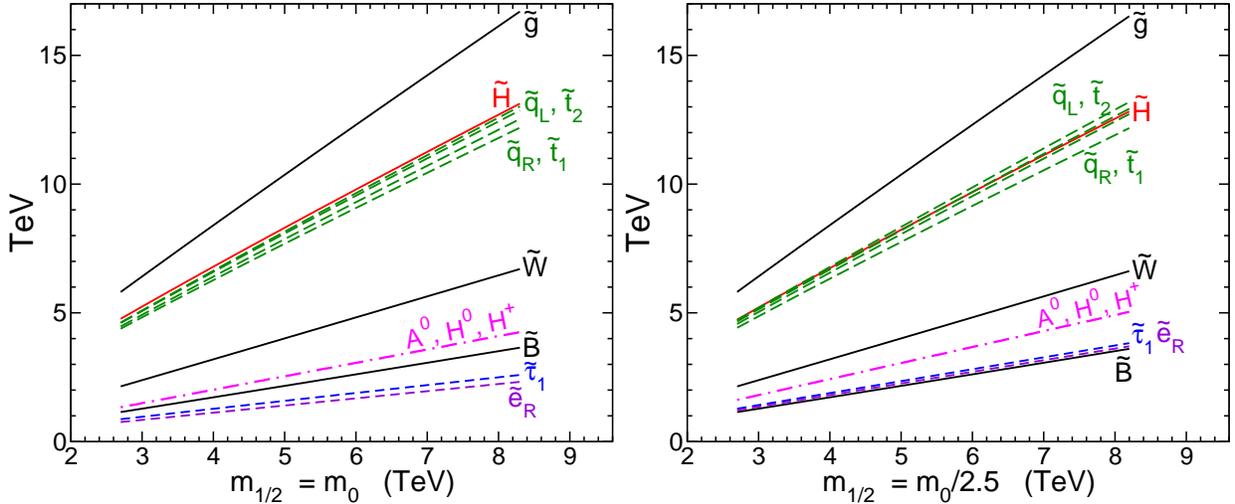

\begin{minipage}[]{0.49\linewidth}
\includegraphics[width=\linewidth,angle=0]{spectrum_U.eps}
\end{minipage}
\begin{minipage}[]{0.49\linewidth}
\begin{flushright}
\includegraphics[width=\linewidth,angle=0]{spectrum_U_heavy.eps}
\end{flushright}
\end{minipage}
\begin{minipage}[]{0.96\linewidth}
\begin{center}
\caption{\label{fig:spectrum_U}
The spectrum of physical masses of the gluino, squarks, sleptons, 
electroweakinos, 
and the heavier Higgs bosons, for a model line with varying GUT-scale
input parameter $m_{1/2}$, with calculated
$M_h$ between 123 GeV (lower edge of $m_{1/2}$) and 127 GeV (higher edge). 
The scalar trilinear coupling parameter $A_0$
and the Higgsino mass parameter $\mu$ are determined by requiring 
electroweak symmetry breaking with $m_Z = 91$ GeV and $\tan\beta=15$.
In the left panel, the common scalar mass is 
$m_0 = m_{1/2}$, and the LSP is a slepton. In the right panel, 
$m_0= 2.5 m_{1/2}$, and the LSP is a bino-like neutralino.
In both cases, $b = m_{1/2}^2$ is imposed at the GUT scale.
The solid lines are, from top to bottom, the gluino, Higgsino, wino, and bino. 
The long-dashed lines are squarks, the short-dashed lines are sleptons, and
the dot-dashed lines are the pseudo-scalar Higgs boson $A^0$
and the nearly degenerate charged and heavy neutral Higgs scalar bosons $H^0, H^\pm$.
}
\end{center}
\end{minipage}
\end{figure}

Figure \ref{fig:spectrum_U} shows the spectrum of the physical masses of selected
superpartners and the heavier
Higgs bosons as a function of the universal gaugino mass $m_{1/2}$ at $Q=M_*$, for the two
choices $m_0 = m_{1/2}$ and $m_0 = 2.5 m_{1/2}$, with $b = m_{1/2}^2$ in both cases. 
The range of $m_{1/2}$ on the horizontal axis shown for this spectrum
corresponds to the estimated $\mbox{123 GeV}<M_h<\mbox{127 GeV}$. 
The general features to be observed include: 
\begin{itemize}
\item the gluino is the heaviest superpartner, with a mass of at least about 6 TeV,
\item the squarks are also too heavy to produce with significant rates 
at the LHC unless there is a major beam energy upgrade,
\item the $\mu$-term is predicted to be very large, resulting in very heavy Higgsino-like charginos and neutralinos also beyond the LHC reach,
\item the heavier Higgs scalar bosons $A^0$, $H^0$, and
$H^\pm$ are nearly degenerate and have masses well in excess of a TeV, and
\item the lightest stau is slightly heavier than the right-handed selectron and smuon,
which escape being the LSP only if $m_0 \gsim 2.5 m_{1/2}$, as noted above.
\end{itemize}

\newpage

\subsection{Optimistic case: non-unified gaugino mass boundary conditions
\label{subsec:nonunified}}

\baselineskip=16.8pt

The model line with unified gaugino masses considered above 
paints a rather bleak picture for the prospects of 
discovering anything new at the LHC. The reason for this is that the 
model predicts top-squark mixing that is not very large, so that 
accommodating $M_h$ near 125 GeV requires large $M_3$, because this is the dominant 
parametric source for the necessary large top-squark masses. 

A much lighter superpartner mass 
spectrum, with hopes for LHC discovery,
can be achieved if one instead considers non-universal gaugino masses. 
By taking $M_3 < M_2$ at the input scale, the quasi-fixed point 
trajectories given approximately by eq.~(\ref{eq:qfpm2Huhat}), and similar results for
$\fixwidehat m^2_{H_d}$ and $b$, provide for correct electroweak 
symmetry breaking with $a_t$ negative and larger in magnitude compared 
to $M_3$. This in turn provides for large top-squark mixing, so that 
$M_h$ can be close to 125 GeV consistent with relatively much lighter 
squarks (and gluino) than found in the previous section with gaugino 
mass unification.

\begin{figure}[!tb]
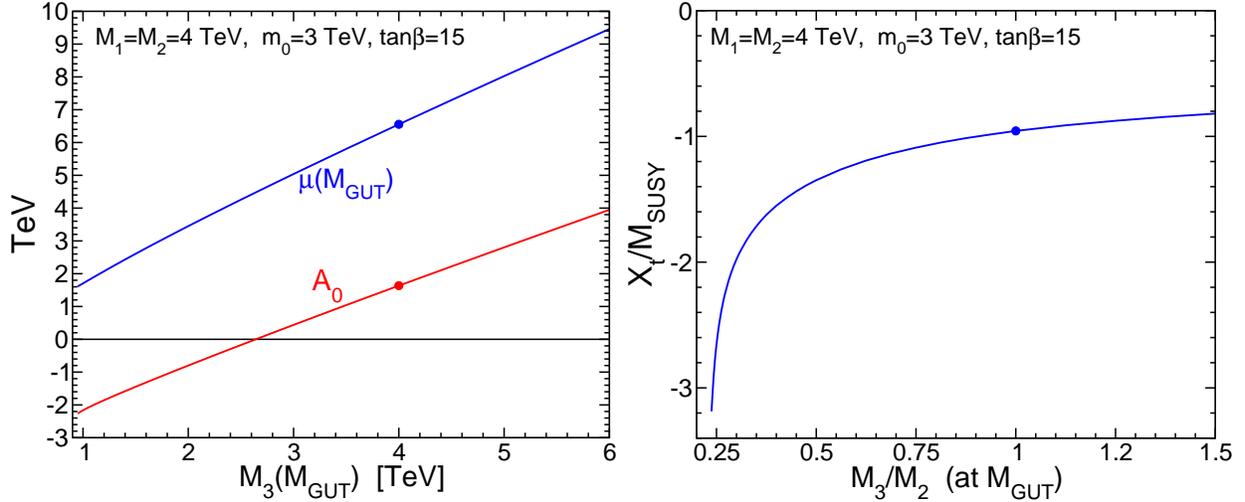

\begin{minipage}[]{0.49\linewidth}
\includegraphics[width=\linewidth,angle=0]{Amu.eps}
\end{minipage}
\begin{minipage}[]{0.49\linewidth}
\begin{flushright}
\includegraphics[width=\linewidth,angle=0]{Xt.eps}
\end{flushright}
\end{minipage}
\begin{center}
\begin{minipage}[]{0.95\linewidth}
\caption{\label{fig:AmuXt}
Results obtained by solving the electroweak symmetry breaking conditions
$m_Z = 91$ GeV and $\tan\beta=15$, for models with fixed GUT-scale parameters
$M_1 = M_2 = 4$ TeV, 
$m_0 = 3$ TeV, and $b = \mbox{(2 TeV)}^2$, as a function of varying $M_3$.
The left panel shows the 
solutions for the GUT-scale parameters $A_0$ and $\mu(M_{\rm GUT})$. The right panel
shows the resulting top-squark mixing parameter
$X_t/M_{\rm SUSY}$ at $Q = M_{\rm SUSY}$, where
$X_t =  a_t/y_t - \mu/\tan\beta$ and 
$M_{\rm SUSY} = \sqrt{m_{\tilde t_1} m_{\tilde t_2}}$. 
The dots are the special case of gaugino mass unification.}
\end{minipage}
\end{center}
\end{figure}

To illustrate this, consider the solutions for $A_0$ and $\mu$
obtained by varying $M_3$ while keeping fixed the input values $M_1=M_2 = 4$ TeV, and 
all non-holomorphic scalar squared masses fixed at $m_0 = 3$ TeV, with 
$b = \mbox{(2 TeV)}^2$ and $\tan\beta=15$. These are shown in the left panel of
Figure \ref{fig:AmuXt}. The trend is for
both $\mu$ and $A_0$ to decrease as $M_3$ is reduced. This implies that 
top-squark mixing is stronger for smaller $M_3/M_2$. 
In the right panel of Figure \ref{fig:AmuXt}, the ratio
$X_t/M_{\rm SUSY}$ at $Q=M_{\rm SUSY}$ 
is shown as a function of $M_3/M_2$ at $Q=M_*$. Here 
$X_t = a_t/y_t - \mu/\tan\beta$ is a top-squark mixing parameter, 
and $M_{\rm SUSY}$ is the geometric mean of the
top-squark masses. It is well-known that $|X_t|/M_{\rm SUSY} \sim \sqrt{6}$ 
tends to approximately maximize $M_h$ for a given $M_{\rm SUSY}$. As can 
therefore be inferred from Figure \ref{fig:AmuXt}b, 
larger $M_h$ will ensue for $M_3/M_2 < 1$, with a particularly interesting range
being about $0.25$ to $0.5$ for this 
ratio.
\footnote{Non-universal gaugino masses, obtained for example 
if the $F$-term that breaks supersymmetry is a singlet under the
Standard Model gauge group but transforms non-trivially under the GUT group
as in refs.~\cite{SU5nonuniversal1}-\cite{SO10nonuniversal2},
have also been used 
to address the supersymmetric little hierarchy problem. 
Coincidentally, naturalness in such
models also prefers $M_3/M_2 \sim 0.3$ at the GUT scale
(see for example refs.~\cite{KaneKing1}-\cite{Yanagida:2013ah}), 
but for a quite different reason, as they work by
making $m^2_{H_u}$ and $|\mu|^2$ individually small.
These models are continuously connected \cite{Younkin:2012ui} in parameter space to the
focus-point scenario \cite{focuspoint}.}
The result is that for a fixed $M_2$ 
and varying $M_3$, the prediction for $M_h$ tends, somewhat coincidentally, 
to be surprisingly not too sensitive to $M_3$. This can be understood as due to 
larger $M_3$ providing for larger logarithmic
contributions to $M_h$ from the overall magnitude of the top-squark mass scale, while smaller $M_3$ yields larger top-squark mixing contributions to $M_h$.

\begin{figure}[!tb]
\begin{minipage}[]{0.49\linewidth}
\includegraphics[width=\linewidth,angle=0]{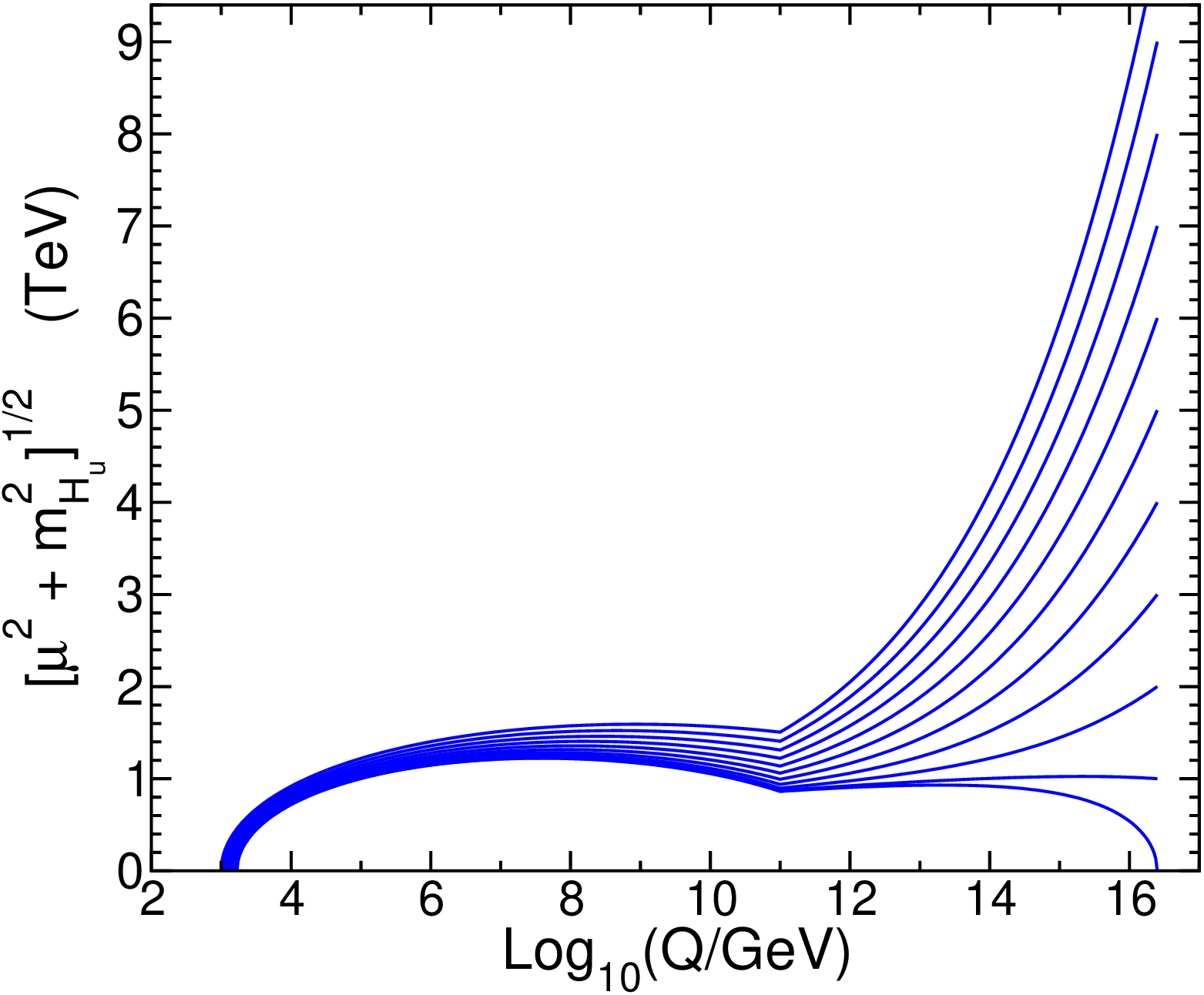}
\end{minipage}
\begin{minipage}[]{0.49\linewidth}
\begin{flushright}
\includegraphics[width=\linewidth,angle=0]{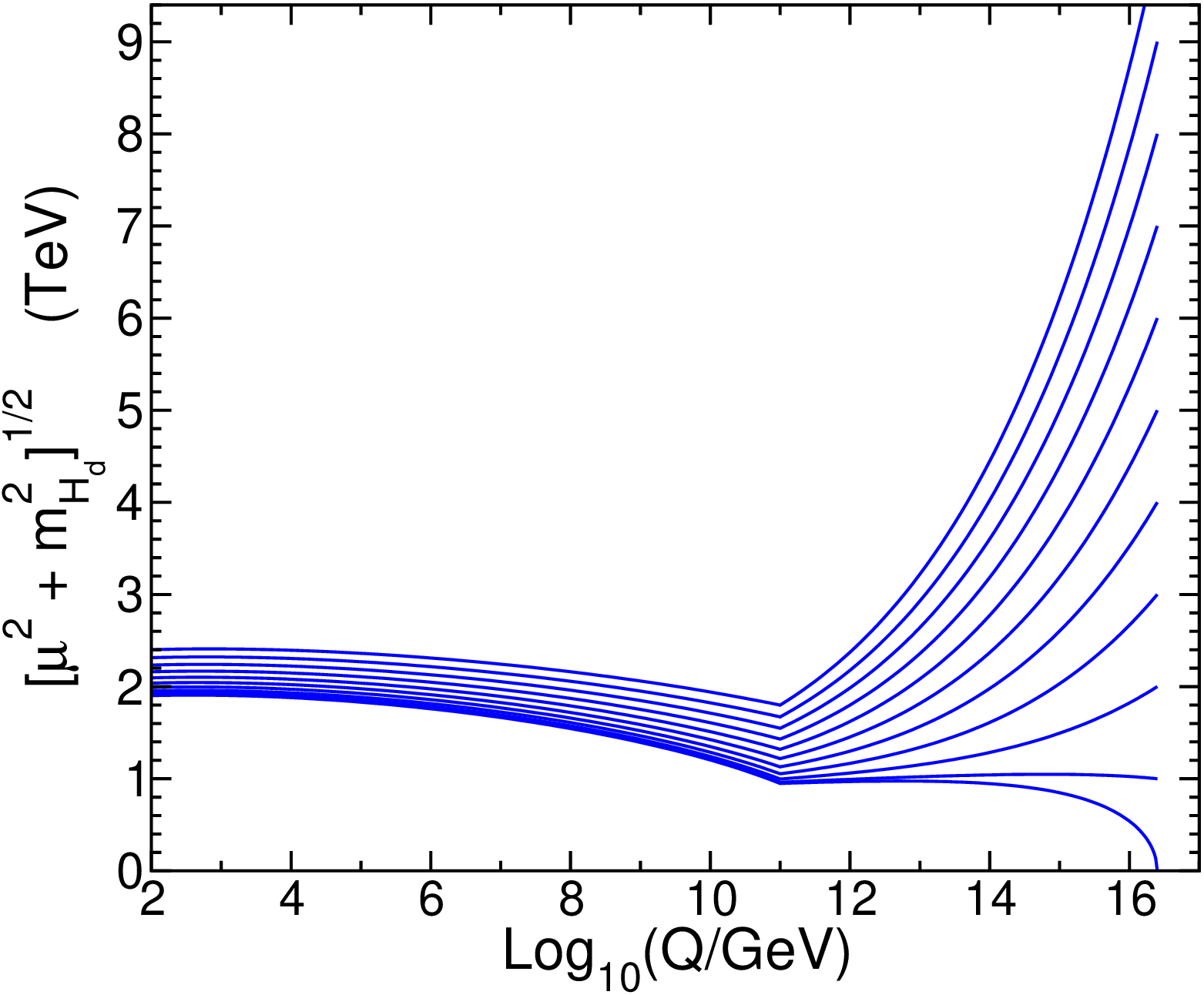}
\end{flushright}
\end{minipage}
\begin{minipage}[]{0.49\linewidth}
\includegraphics[width=\linewidth,angle=0]{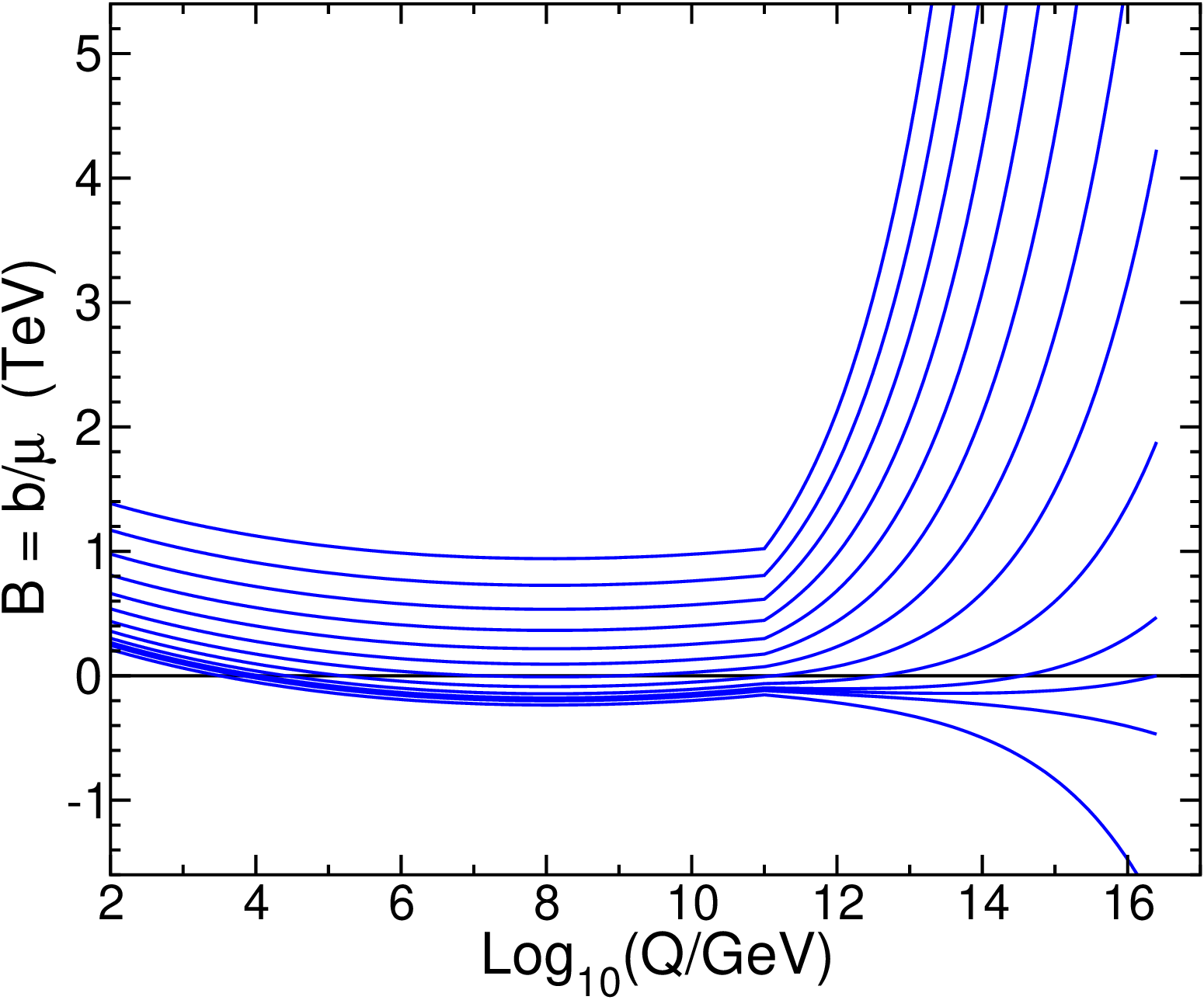}
\end{minipage}
\hspace{0.04\linewidth}
\begin{minipage}[]{0.45\linewidth}
\caption{\label{fig:runQ_NONU_Higgs}
Renormalization group running of the Higgs mass parameters 
$(m^2_{H_u} + \mu^2)^{1/2}$, 
$(m^2_{H_d} + \mu^2)^{1/2}$, and 
$B = b/\mu$,
for a model with non-universal gaugino masses 
$M_{1} = 2000$ GeV, $M_{2} 
= 4000$ GeV, and $M_3 = 1200$ GeV at the GUT scale. Each line is the 
RG trajectory for a different boundary condition of 
the common scalar squared mass $m_0^2 = b$
at the GUT scale, showing the approach to the 
quasi-fixed point behavior at low RG scales.}
\end{minipage}
%

\vspace{-0.3cm}

\begin{center}
\noindent\rule{1.01\linewidth}{0.8pt}
\end{center}

\begin{minipage}[]{0.49\linewidth}
\includegraphics[width=\linewidth,angle=0]{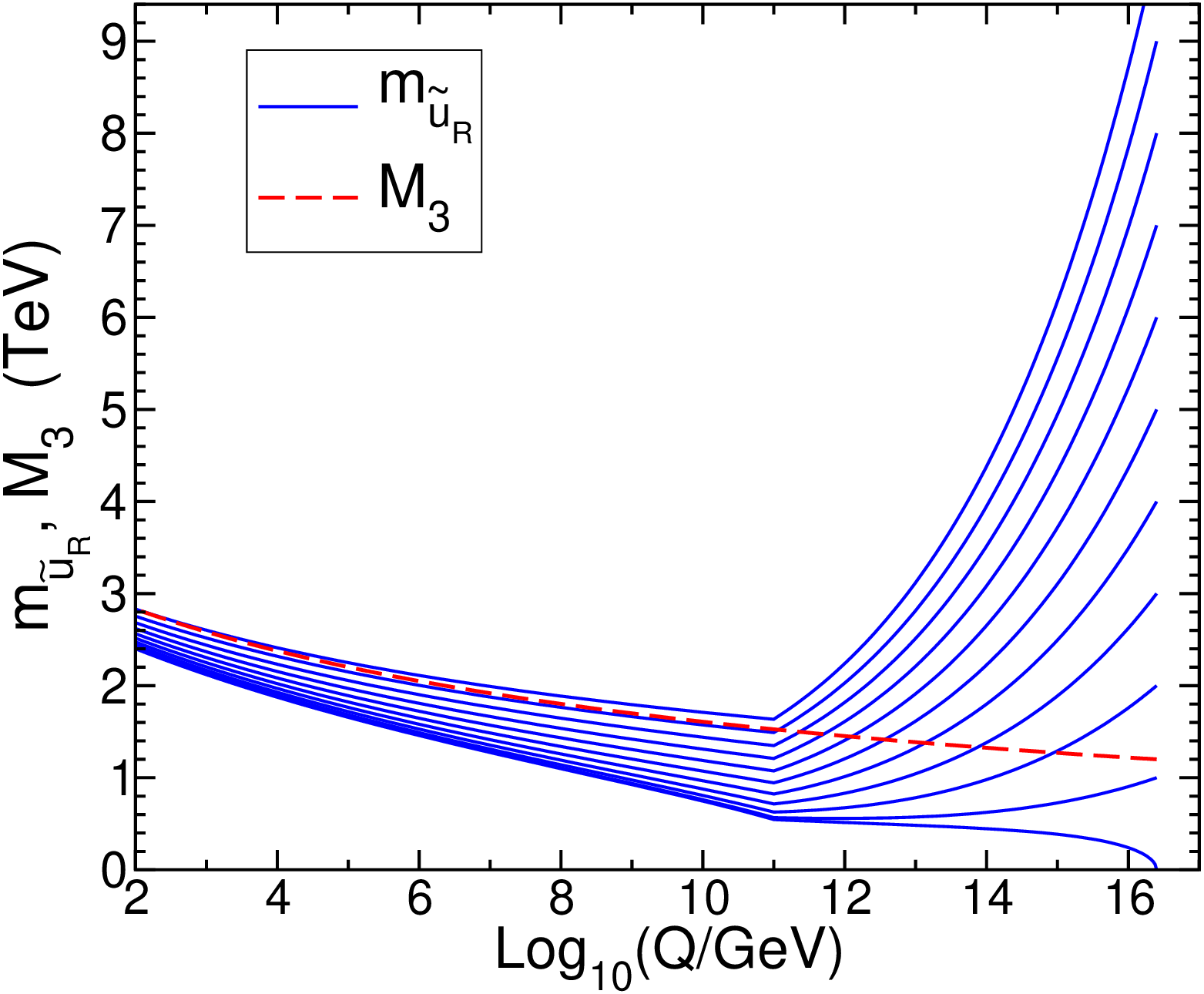}
\end{minipage}
\begin{minipage}[]{0.49\linewidth}
\begin{flushright}
\includegraphics[width=\linewidth,angle=0]{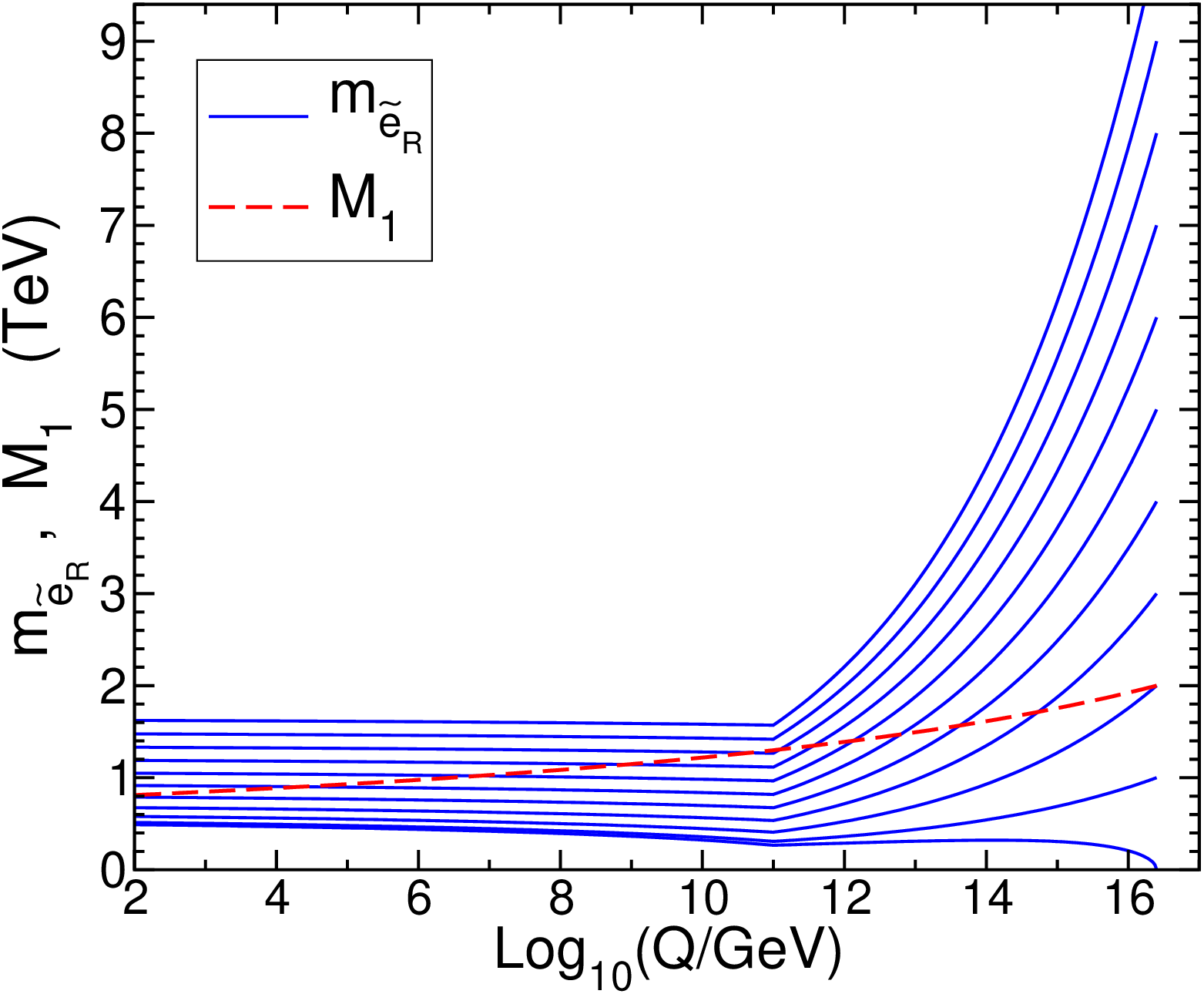}
\end{flushright}
\end{minipage}
\caption{\label{fig:runQ_NONU_sfermions}
RG running of the squark and slepton mass parameters 
$m_{\tilde u_R}$ (left panel) and $m_{\tilde e_R}$ (right panel) 
for a model with non-universal gaugino masses
$M_{1} = 2000$ GeV, $M_{2} = 4000$ GeV, and $M_3 = 1200$ GeV 
at the GUT scale. 
Each line is the RG trajectory for a 
different boundary condition of the common scalar mass at the GUT 
scale, showing the approach to the quasi-fixed point 
behavior. The running gluino and bino masses $M_3$ and $M_1$ are 
also shown for comparison.}
\end{figure}

Consider an example model line 
defined by gaugino masses chosen at $M_*$ to be 
$M_3 = 1200$ GeV,
$M_2 = 4000$ GeV, and
$M_1 = 2000$ GeV, and varying universal $m_0^2=b$, with $A_0$ and $\mu$
determined by correct electroweak symmetry breaking
with $m_Z = 91$ GeV and $\tan\beta = 15$. This choice of parameters 
is made because it
results in $M_h$ close to 125 GeV (with some mild dependence on $m_0$ that is well within 
the theoretical and parametric uncertainties).
The resulting RG trajectories indicating the attraction to the quasi-fixed points
are shown in Figure \ref{fig:runQ_NONU_Higgs} for 
$\fixwidehat m_{H_u}^2$, $\fixwidehat m_{H_d}^2$, and $B$. 
In comparison to the universal gaugino mass case, the results for 
$\fixwidehat m_{H_u}^2$ at scales at and below $Q = \Lambda$ are 
considerably smaller, so that there is arguably less fine-tuning 
involved to obtain electroweak symmetry breaking with the observed 
$m_Z$. For example, for all $m_0^2 < (4$ TeV$)^2$, the value of $\fixwidehat 
m_{H_u}$ at $Q=\Lambda$ is less than 1 TeV, and it maintains a more moderate 
slope throughout its running, compared to the scenario with unified gaugino masses
at the GUT scale.

The running mass of a right-handed squark is shown in the left panel of 
Figure \ref{fig:runQ_NONU_sfermions}, along with the gluino running mass 
parameter $M_3$. Both the gluino and squarks can be much lighter than in 
the universal gaugino mass case, and can easily be less than 3 TeV, and therefore
accessible to the LHC with sufficient integrated luminosity. No 
effort was made to fine-tune the chosen model to be optimized in this regard, 
so that even somewhat lighter squarks and gluino are possible. As in the 
universal gaugino mass case, the right-handed squarks are lighter than 
the gluino, but the left-handed squarks can be heavier than the gluino, 
due to the RG influence of
the much larger value of $M_2^2$ compared to $M_3^2$ here. 

The right panel of Figure 
\ref{fig:runQ_NONU_sfermions} shows the running of the right-handed 
selectron mass compared to the running bino mass $M_1$. This is qualitatively 
quite similar to the situation in the universal gaugino mass case. 
However, one important difference is that in this case the lighter stau 
(not shown in Figure \ref{fig:runQ_NONU_sfermions}) 
can have a slightly smaller mass than the right-handed 
selectron and smuon. As before, the LSP is predicted to be the bino-like 
neutralino if $m_0 \gsim 2.5 M_1$ at $Q=M_*$; otherwise it is a 
stau, and $R$-parity violation can be invoked to avoid a disastrous stable
charged relic from the early universe.

\begin{figure}
\begin{minipage}[]{0.57\linewidth}
\begin{flushright}
\includegraphics[width=\linewidth,angle=0]{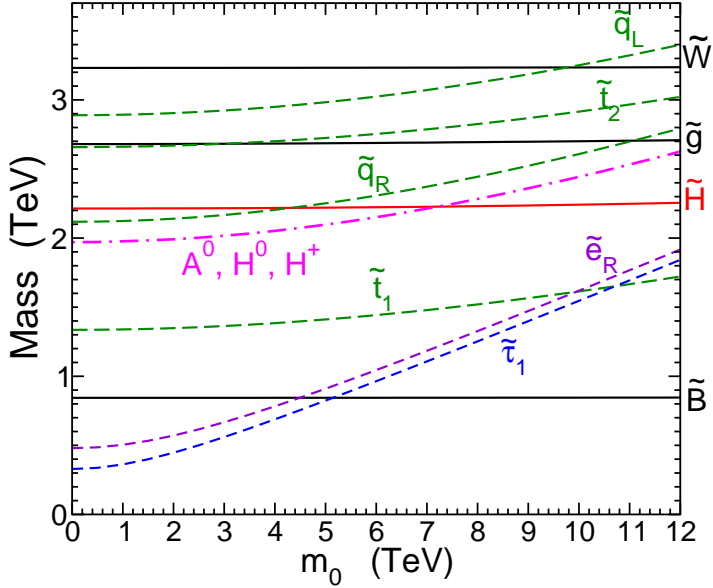}
\end{flushright}
\end{minipage}
\hspace{0.02\linewidth}
\begin{minipage}[]{0.39\linewidth}
\caption{\label{fig:spectrum_NONU}
The spectrum of physical masses of the gluino, squarks, sleptons, 
electroweakinos, 
and the heavier Higgs scalar bosons, for a model line with fixed GUT-scale
input parameters $M_3 = 1200$ GeV, $M_2 = 4000$ GeV, $M_1 = 2000$ GeV,
and $b = (\mbox{2000 GeV})^2$,
as a function of $m_0$.
The scalar trilinear coupling parameter $A_0$
and the Higgsino mass parameter $\mu$ are determined by requiring 
electroweak symmetry breaking with $m_Z = 91$ GeV and $\tan\beta=15$. 
The lightest neutral Higgs boson mass is close to 125 GeV, 
due to large top-squark mixing.}
\end{minipage}
\end{figure}

Figure \ref{fig:spectrum_NONU} shows the physical mass spectrum as a 
function of $m_0$.
Some salient features of this model line with $M_3/M_2 \sim 0.3$ 
that are different from the 
universal gaugino mass case of the previous subsection are:
\begin{itemize}
\item The superpartner mass spectrum is relatively 
compressed compared to scenarios based on unified gaugino masses at the GUT scale, 
but (in this example, at least) not enough to dramatically 
impact typical LHC search strategies for a given gluino mass, 
\item The heaviest superpartner is a wino-like chargino/neutralino
(or perhaps a left-handed squark if $m_0$ is very large), 
so that decays through (off-shell) winos are suppressed,
\item The gluino and squarks are beyond the reach of the collected LHC data
as of this writing, but might be
seen at a future luminosity or higher energy upgrade, 
\item The lightest of the sleptons is a stau, rather than a 
right-handed selectron or smuon.
\end{itemize}

\newpage

\section{Outlook\label{sec:outlook}}
\setcounter{equation}{0}
\setcounter{figure}{0}
\setcounter{table}{0}
\setcounter{footnote}{1}

\baselineskip=17.2pt

In this paper, I have noted that in scalar sequestering models of 
low-energy supersymmetry, the interplay between the strong superconformal 
dynamics in the hidden sector and the perturbative renormalization effects 
of the visible sector results in running towards non-trivial 
quasi-fixed point RG trajectories at an intermediate scale. The scalar squared masses
of the theory are suppressed, but run to values that are 
non-negligible, because the scaling parameter $\Gamma$ cannot be too large.
It is tantalizing that the combination $m^2_{H_u} + |\mu|^2$ is 
subject to this suppression, because this is the combination that should be small 
at the TeV scale
in order to solve the little hierarchy problem. In the end, taking into account
the constraints on anomalous dimensions, it appears that there is  
some amelioration of the little hierarchy problem, due to the suppression
proportional to $3/8 \pi^2 \Gamma$ in eq.~(\ref{eq:qfpm2Huhat}), especially
in models with $M_3 < M_2$ at the high input scale.
However, I would not argue that this attains a completely compelling solution
to the problem (which is in any case somewhat subjective), 
as the experimental value of 
$m_Z^2$ is still quite small compared to the individual MSSM parameter 
contributions to it. 

Models of this kind do at least
have the virtue of providing some predictive power, compared to the general MSSM, 
due to the renormalization group quasi-fixed point structure. The superpartner 
mass spectrum depends only weakly, and in specific calculable ways, 
on the input values of the dimension two parameters. 
It should be remarked that the details of the scalar 
sequestering necessarily cannot be predicted with fine precision, given 
the lack of knowledge of the hidden sector dynamics. The anomalous dimensions of a 
putative superconformal theory are bounded but not known, and the hidden sector is likely
to be approximately, but not exactly, superconformal. While the predictivity
of this model framework is thus limited, general and
qualitative statements can still be made.
A prominent example of this predictivity is that imposing gaugino mass unification seems
to require superpartners definitely  
beyond the reach of the LHC even after a high-luminosity upgrade, 
given the observed mass of the lightest Higgs scalar boson at $M_h = 125$ GeV.
By relaxing the assumption to allow $M_3 < M_2$ at the input scale, I argued that the
scenario becomes subjectively more natural, while also 
allowing for a chance at discovery in the future at the LHC.

{\it Acknowledgments:} I thank Simon Knapen and Martin Schmaltz for useful conversations.
This work was performed in part in the exalted atmosphere of
the Aspen Center for Physics, which is supported by National Science Foundation
grant PHY-1607611.
This work was supported in part by the 
National Science Foundation grant number PHY-1719273. 

\vspace{-0.25cm}
 

\end{document}